\documentclass[]{aa}  

\usepackage{graphicx}
\usepackage{txfonts}
\usepackage{lipsum}
\usepackage{subcaption}         
\usepackage{lscape}             
\usepackage{placeins}           
                                
\usepackage{multirow}%
\usepackage{amsmath,amssymb,amsfonts}%
\usepackage{mathrsfs}%
\usepackage[title]{appendix}%
\usepackage{xcolor}%
\usepackage{textcomp}%
\usepackage{manyfoot}%
\usepackage{booktabs}%
\usepackage{algorithm}%
\usepackage{algorithmicx}%
\usepackage{algpseudocode}%
\usepackage{listings}%

\usepackage{xpatch}
\usepackage{hyperref}
\hypersetup{
	colorlinks=true,
	breaklinks=true,
	citecolor=blue,
	allcolors=blue,
	frenchlinks=true
}

\makeatletter
\xpatchcmd\NAT@citex
{%
	\@citea\NAT@hyper@{%
		\NAT@nmfmt{\NAT@nm}%
		\hyper@natlinkbreak{\NAT@aysep\NAT@spacechar}{\@citeb\@extra@b@citeb}%
		\NAT@date
	}%
}
{%
	\@citea
	\NAT@nmfmt{\NAT@nm}%
	\NAT@aysep\NAT@spacechar
	\NAT@hyper@{\NAT@date}%
}
{}{}
\xpatchcmd\NAT@citex
{%
	\@citea\NAT@hyper@{%
		\NAT@nmfmt{\NAT@nm}%
		\hyper@natlinkbreak{\NAT@spacechar\NAT@@open\if*#1*\else#1\NAT@spacechar\fi}%
		{\@citeb\@extra@b@citeb}%
		\NAT@date
	}%
}
{
	\@citea
	\NAT@nmfmt{\NAT@nm}%
	\NAT@spacechar\NAT@@open\if*#1*\else#1\NAT@spacechar\fi
	\NAT@hyper@{\NAT@date}%
}
{}{}
\makeatother

\makeatletter
\renewcommand*\aa@pageof{, page \thepage{} of \pageref*{LastPage}}
\makeatother

\begin{document}

   \title{The supersonic nature of jellyfish galaxies}

%

   \author{
    Alessandro Ignesti\inst{1}
    \and
    Francesca Loi\inst{2}
    \and
    Antonino Marasco\inst{1}
    \and
    Benedetta Vulcani\inst{1}
    \and
    Bianca M. Poggianti\inst{1}
    \and
    Christoph Pfrommer\inst{3}
    \and
    Marco Gullieuszik\inst{1}
    \and
    Alessia Moretti\inst{1}
    \and
    Paolo Serra\inst{2}
    \and
    Stephanie Tonnesen\inst{4}
    \and
    Rory Smith\inst{5,6}
    \and
    Cecilia Bacchini\inst{7}
    \and
    Marc A. W. Verheijen\inst{8}
    \and
    Myriam Gitti\inst{9,12}
    \and
    Anna Wolter\inst{10}
    \and
    Koshy George\inst{11}
    \and
    Yara Jaffe\inst{5,6}
    \and
    Rosita Paladino\inst{12}
    \and
    Giorgia Peluso\inst{13}
    \and
    Mario Radovich\inst{1}
    \and
    Augusto E. Lassen\inst{1}
    \and
    Neven Tomi\v{c}i\'{c}\inst{14}
    \and
    Peter Kamphuis\inst{15}
}

\institute{
    INAF - Osservatorio Astronomico di Padova, Vicolo dell'Osservatorio 5, 35122 Padova (PD), Italy \label{inst1}\\\email{alessandro.ignesti@inaf.it}
    \and
    INAF - Osservatorio Astronomico di Cagliari, Via della Scienza 5, 09047 Selargius (CA), Italy \label{inst2}
    \and
    Leibniz Institute for Astrophysics Potsdam (AIP), An der Sternwarte 16, 14482 Potsdam, Germany \label{inst3}
    \and
    Center for Computational Astrophysics, Flatiron Institute, 162 5th Avenue, 10010 New York (NY), USA \label{inst4}
    \and
    Departamento de F\'isica, Universidad T\'ecnica Federico Santa Mar\'ia, Avenida Espa\~na 1680, Valpara\'iso, Chile \label{inst5}
    \and
    Millennium Nucleus for Galaxies (MINGAL), Valpara\'iso, Chile \label{inst6}
    \and
    DARK, Niels Bohr Institute, University of Copenhagen, Jagtvej 155, 2200 Copenhagen, Denmark \label{inst7}
    \and
    Kapteyn Astronomical Institute, University of Groningen, Landleven 12, 9747 AD Groningen, Netherlands \label{inst8}
    \and
    Department of Physics and Astronomy "Augusto Righi", University of Bologna, Via Gobetti 93/2, 40129 Bologna (BO), Italy \label{inst9}
    \and
    INAF - Osservatorio Astronomico di Brera, Via Brera 28, 20121 Milano (MI), Italy \label{inst10}
    \and
    University Observatory, LMU Faculty of Physics, Scheinerstrassee 1, 81679 M\"unchen, Germany \label{inst11}
    \and
    INAF - Istituto di Radioastronomia, Via Piero Gobetti 101, 40129 Bologna (BO), Italy \label{inst12}
    \and
    INAF - Osservatorio di Astrofisica e Scienza dello Spazio Bologna, Via Piero Gobetti, 93/3, 40129 Bologna (BO), Italy \label{inst13}
    \and
    Department of Physics, Faculty of Science, University of Zagreb, Bijeni\v{c}ka 32, 10 000 Zagreb, Croatia \label{inst14}
    \and
    Astronomical Institute (AIRUB), Ruhr-University Bochum, Faculty of Physics and Astronomy, 44780 Bochum, Germany \label{inst15}
}
\authorrunning{Ignesti et al.}

   \date{Received September 30, 20XX}

 
  \abstract
{All gas-rich galaxies in cluster environments are expected to experience ram-pressure stripping from the intra-cluster medium. However, only a fraction of these develop ongoing star-formation in their stripped tail, becoming the so-called ``jellyfish'' galaxies. 
In this work we provide observational evidence that magnetic fields can signal differences in the extraplanar star formation and explore what are the physical conditions that lead to the formation of a jellyfish galaxy. We first focus on JO147, a jellyfish galaxy that features weak star formation activity in its tail. Using MeerKAT radio continuum observations, we discover polarized emission only in a small fraction of its tail, with an average fraction of $~10\%$, and a low Mach number $\mathcal{M}=1.3-1.6$, which suggests a possible association between magnetic field draping, shock-compression of the gas, and extraplanar star formation activity. Then, we test this scenario in a sample of 17 jellyfish galaxies from the GASP project. We combine dynamical models for their orbits within the host clusters with realistic cluster temperature profiles to infer their Mach number, and we find a positive correlation between it and the star formation activity in their tail. We conclude that  supersonic motion is a necessary condition for triggering star formation in the stripped tails of jellyfish galaxies. Our findings provide empirical evidence that the critical factor preventing the stripped gas evaporation is the shock compression induced by the supersonic motion through the cluster. This process likely enhances the magnetic field surrounding the galaxy and the properties of the stripped material. }

   \maketitle

\nolinenumbers
\section{Introduction}
\label{intro}
Galaxies falling into galaxy clusters are subject to an external ram pressure resulting from their large speed relative to the intracluster medium (ICM) \citep[][]{Gunn1972}, the hot plasma filling the cluster volume. Ram pressure affects the galaxy's interstellar medium (ISM) and circumgalactic medium (CGM). In galaxies moving at several hundred km s$^{-1}$ relative to the ICM, it can overcome the gravitational binding of the stellar disk, stripping gas from the disk \citep[][]{Hester2006, Smith2010, Poggianti2016, Boselli2022} and the halo \citep[][]{Sparre2024_CGM} of the infalling galaxy. The stripped ISM, with typical temperature of $10^{2-5}$ K, \citep[][]{Spitzer_1978} can interact with the ICM, which is a weakly magnetized plasma characterized by a density of $10^{-4}-10^{-3}$ particles cm$^{-3}$, a temperature of $10^{7-8}$ K \citep[][]{Sarazin1988}, and $\mu$G-level magnetic fields \citep[][]{Govoni2004}. The large temperature and velocity differences between the two phases imply short evaporation timescales for the stripped ISM \citep[${\sim}10^{7-8}$ yr,][]{Klein1994, Vollmer_2001} due to a combination of hydrodynamical instabilities and thermal conduction. As the typical stripping timescales are an order of magnitude larger \citep[$\sim$10$^{8-9}$ yr,][]{Smith2022b,Rohr_2023}, the expectation is that the stripped ISM will completely evaporate in the ICM during this process. Yet, in the so-called jellyfish galaxies \citep[][]{Ebeling_2014,Waldron2023,Poggianti2025}, we observe trails of stripped ISM extending for tens of kpc and hosting active star-forming regions. Similarly, clouds of warm and neutral gas has been observed at hundreds of kpc from their original hosting galaxy \citep[][]{Serra_2024,Sun_2025}. These results proves the existence of a mechanism that can stabilize the stripped ISM, thus extending its survival outside of the stellar disk and permitting it to cool down and form new stars.\\ 

Jellyfish galaxy tails derive from stripped ISM, which mixes with the warm CGM and the hot ICM wind. To become thermally unstable and, thereby potentially leading to star formation \citep{Lee2022,Sparre2024_CGM}, the mixed material requires a cooling rate that exceeds the growth rate of the Kelvin–Helmholtz instability, which would otherwise disrupt and dissolve the tail \citep{Gronke2018,Sparre2019,Sparre2020,Li2020}. The presence of ordered magnetic fields along the ICM-cold gas interface can significantly modify the stripping process and star formation in tails due to suppressing thermal conduction between the hot and cold phases, stabilizing against hydrodynamical instabilities \citep[][]{Frank_1996,McCourt_2015,Sparre2020,Sparre2024}, as well as reducing the gas mass loss \citep[][]{Rintoul_2025}. The presence of magnetic fields in the stripped material is naturally expected due to both internal and external factors. On the one hand, the stripped material is expected to be magnetized because it contains the ISM magnetic field bound to the thermal gas being removed by the ram pressure \citep[][]{Vollmer_2004,Ignesti2023, Vollmer_2024,2024MNRAS.533.1394M}. In this scenario, due to the turbulent, small-scale motions in the stripped material \citep[][]{Li_2023,Ignesti_2024} the stripped tail is expected to show a low degree of polarized synchrotron emission, as consequence of the magnetic field disordered structure and the strong Faraday depolarization resulting from the thermal plasma mixed with the nonthermal components. 

On the other hand, the weak magnetic field permeating the ICM can accumulate around the infalling galaxy via the so-called magnetic draping \citep[][]{Dursi2008,Pfrommer2010}, which would naturally provide a way for jellyfish galaxies to surround themselves with large-scale magnetic fields accreted from the environment. Numerical simulations \citep[][]{Dursi2008} indicate that a prerequisite for the formation of the large-scale, ordered magnetic drape is that the galaxy's velocity must exceed the local Alfv\'en speed $V_A=B/\sqrt{4\pi\rho}$, where $B$ is the magnetic field and $\rho$ is the ion mass density, to bend the external magnetic field on its surface. This condition is virtually always satisfied in galaxy clusters, where the typical ICM Alfv\'en speed is of the order of several tens of km s$^{-1}$ and the cluster velocity dispersion is typically of the order of several hundreds of km s$^{-1}$ \citep[][]{Girardi_1993}. Furthermore, in the case of supersonic motion, the ICM magnetic field can be significantly amplified at the curved bow-shock propagating into an inhomogeneous ICM, which adiabatically enhances the ICM magnetic field via shock compression and injects turbulence that could further amplify the magnetic field via a small-scale dynamo. In fact, for galaxies moving supersonically in the ICM, numerical simulations predict the formation of an ordered magnetic drape extending for tens of kpc in the galaxy wake \citep[][]{Sparre2020,Sparre2024}. This mechanism would be the one responsible for the formation of an ordered field ``shielding'' the stripped material. In this scenario, the galaxy is also expected to show a high degree of polarized emission thanks to the magnetic field being ordered on large scales and the fact that it would be less affected by the Faraday depolarization induced by the stripped material. As in galaxy clusters the speed of sound is comparable to the cluster velocity dispersion, supersonic draping is expected to be at work for jellyfish galaxies, which typically are the fastest cluster galaxies \citep[][]{Jaffe_2018}, providing a potential explanation for the origin of the long star-forming tails.\\

The evidence that jellyfish galaxies can form a strong magnetic drape on their contact surface with the ICM has been obtained thanks to deep radio continuum observations of the jellyfish galaxy JO206 \citep[][]{Mller2021}. Highly polarized emission was detected both in the stellar disk and along the stripped ISM traced by H$\alpha$ emission. The high degree of polarization indicates that the radio emission originated from a magnetized medium located outside of the stripped ISM, which would have otherwise depolarized the emission via Faraday rotation. The polarization angle, which traces the magnetic field topology, indicated that the magnetic field was aligned with the stripped material. This milestone result demonstrated that jellyfish galaxies can be in a condition to form the magnetic field configuration that protects the stripped ISM from the ICM. 

In this work, we extend this study to another galaxy, JO147 (RA 13:26:49.73, DEC -31:23:45.5, $z=0.0506$, also known as SOS 114372) to provide observational evidence that the shock compression resulting from the supersonic galaxy motion is the critical factor in determining the formation of the jellyfish galaxy's star-forming tails. JO147 resides in the galaxy cluster Abell 3558 ($z=0.04889$),  in the central region of the Shapley Supercluster \citep[][]{Shapley_1930}, which is one of the richest and most massive concentrations of gravitationally bound galaxy clusters in the local Universe \citep[e.g.,][ and references therein]{Bardelli1996,Venturi_2000,Rossetti_2007,Merluzzi_2015,Venturi_2022,2024MNRAS.533.1394M,GdG_2025}. The galaxy shows a trail of stripped ISM traced by extended H$\alpha$ emission, similarly to the case of JO206. However, unlike JO206, it hosts a negligible amount of extraplanar star formation \citep[][]{George_2025}. It thus represent the ideal candidate to determine which conditions have not verified that lead to such difference in these galaxies. Here we present the results of new observations at 1.4 GHz of JO147 taken with the MeerKAT radio telescope \citep[][]{Jonas_2016} to map its magnetic field morphology. 

The manuscript is structured as follows. Details about the data calibration and imaging are reported in Section \ref{data_red}, and the results are reported in Section \ref{results}. In Section \ref{turbo} the new results are discussed to investigate how the galaxy motion can influence its radio continuum emission, and the emerging physical framework is further tested and explored in Section \ref{test_MC}. Throughout the paper, we adopt a $\Lambda$CDM cosmology with $\Omega_{\Lambda}=0.7$, $\Omega_{\rm m}=0.3$, and $H_0=70$ km s$^{-1}$ Mpc$^{-1}$. For the clusters analyzed in this work it results in $1''\simeq1$ kpc.

\section{Data processing}
\label{data_red}
JO147 has been observed for eight hours (Project IDs 1671585951, 1672809833, PI M\"uller) to map the nonthermal synchrotron radiation emitted from the cosmic ray electrons previously accelerated in the disk and, later on, displaced by the ram pressure. MeerKAT broadband full-Stokes data were acquired between December 2022 and January 2023 using the 32k correlator during two observing sessions (Project IDs: 20221028-0010 and 20221028-0011). Each session included a 10-minute scan of a primary calibrator (either J0408-6545 or J1939-6342). A secondary calibrator (J1323-4452) was observed for 2 minutes both before and after the target scans. The target source, JO147, was observed for 30 minutes per scan, accumulating a total of 4 hours of on-source integration time per observing session. Additionally, two 5-minute scans of the polarized calibrator (J1331+3030) were conducted at different parallactic angles to optimize sensitivity in the cross-hand correlations, ensuring adequate signal strength for polarization calibration. \\

We performed the data reduction and imaging with the CARACal software \citep[][]{jozsa2020} following the same steps described in \citet[][]{Loi_2025}.
First of all we transfer the data binning in frequency channel of 208 kHz to reduce the data volume obtaining 2 data sets of $\sim$0.9\,TB. 
After splitting the calibrators in a new measurement set, we flag autocorrelations, shadowed antennas, and the frequency
ranges 1379.6--1382.3 MHz and 1420.36--1420.56 MHz affected by the GPS L3 signal and by absorption/emission of neutral hydrogen from the Milky Way respectively. We used the AOFlagger \citep[][]{offringa2012} tool to excise the remaining RFIs using the QUV Stokes visibilities. 
We averaged the data to a frequency resolution of 1 MHz and we derived the calibration terms excluding baselines shorter than 100 m. We adopted a sky model for the primary calibrator and we solved for antenna-based time-independent delays, complex gains, and complex bandpass, applying at each step all the calibration terms derived up to that point. After applying the primary calibrator delay and bandpass to the secondary calibrator, we derived for every scan the antenna-based frequency independent complex gains of the secondary. We scaled the resulting gain amplitudes by bootstrapping the flux scale based on the primary calibrator gains.
We applied the solutions to the calibrators and for the polarized calibrator we assumed the spectro-polarimetric properties reported in the NRAO website\footnote{\url{https://science.nrao.edu/facilities/vla/docs/manuals/obsguide/modes/pol}} and in \citet[][]{perley2017}.
We used the polarized calibrator to solve for the cross-hand delay and phase and the primary calibrator to derive the off-axis leakage term in this order, applying at each step all the calibration terms derived up to that point.
We split the target from the original data sets applying all the calibration terms and limiting the frequency range to 0.9--1.65\,GHz to avoid bandpass rolloffs. We flagged autocorrelations, shadowed antennas, and the frequency ranges 1379.6--1382.3 MHz and 1420.36--1420.56 MHz. We used tricolour\footnote{{\url{https://github.com/ratt-ru/tricolour}}} to excise the remaining RFIs reaching a total flagging percentage of $\sim$50\%.
We averaged the data to a frequency resolution of 1\,MHz and we proceeded with the imaging and self-calibration.\\

We performed the imaging on a regular grid of 6800 pixels with a pixel size of 1.5\,arcsec and Briggs \texttt{robust=-0.5}. We use 4 coefficient to model the spectral shape of the clean components at fixed R.A. and Dec., producing 8 frequency channels in output. After a first blind deconvolution, we derived a mask with the SoFiA--2 \citep[][]{serra2015,Westmeier2021MNRAS.506.3962W} software, using a running window of 50 pixels to accurately evaluate the noise in every part of the image and an initial threshold of 8$\sigma$, where $\sigma$ is the local noise. To improve the masking of discrete sources we enabled the smooth and clip algorithm in SoFiA that iteratively smooths the image with a user-defined set of smoothing kernels, measuring the noise level on each smoothing scale, and adding all pixels with an absolute flux above the 8$\sigma$ threshold. After deriving the mask we repeated the imaging and we proceeded with a first cycle of self-calibration with cubical, solving for the delay terms every 32 seconds.
We then derived a new mask with SoFiA on the last image, using a threshold of 5$\sigma$, and we repeated the imaging on the self-calibrated data using this mask. To improve the results we repeated the self-calibration-imaging loop using a new mask with a 3$\sigma$ threshold.
To derive the polarized intensity we performed the imaging of the Q and U Stokes parameters between 900 MHz and 1.4 GHz (to avoid the off-axis leakage effects). The deconvolution is made by combining all the frequency channels and all the Stokes parameters in quadrature. We produced Q and U output cubes having a frequency channel width of 5 MHz. After a convolution of all the planes of the cubes to a common resolution of 12\,arcsec we run the Rotation Measure Synthesis technique to avoid bandwidth depolarization. We used the RMtools \citep[][]{rmtools} software to explore the Faraday depth between -200 and 200\,rad/m$^2$ with a step of 5\,rad/m², weighting every frequency channels by the inverse of the average noise squared. The polarized intensity is defined as the peak of the Faraday Dispersion Function estimated along the Faraday depth axis, pixel per pixel. Once we estimated the QU noise in the bandwidth-averaged images with SoFiA, we corrected the polarized intensity for the Ricean bias following \citep[][]{george2012}. The polarization shown in Fig. \ref{image} is above a threshold of 4 times the mean noise in the bandwidth-averaged Q and U Stokes images that is 6.5$\mu$Jy/beam. The polarized vectors are proportional to the fractional polarization and the orientation is the de-rotated B-vector. The total intensity contours, drawn at \(3\sigma\) with $\sigma$=7$\mu$Jy/beam, reveal nonthermal synchrotron radiation both on the disk and extending along a 60 kpc tail, consistent with the findings reported by \citep[][]{2024MNRAS.533.1394M}. Both the total and polarized intensity images have a resolution of 12\,arcsec.

\section{Magnetic field orientations}
\subsection{New evidence of extraplanar polarized emission in JO147}
\label{results}
\begin{figure}[t!]
    \centering
    \includegraphics[width=\linewidth]{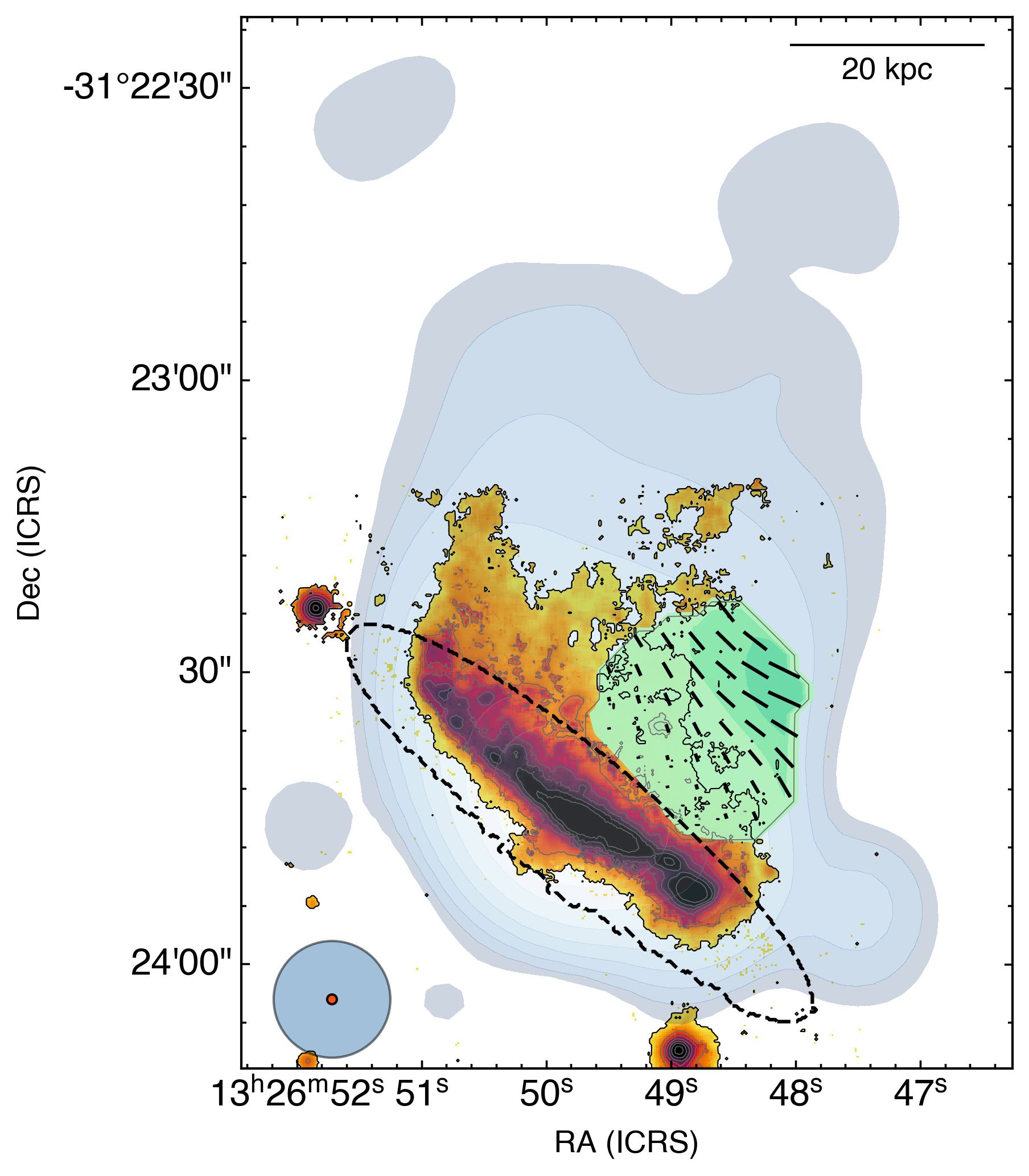}
    \caption{Composite MUSE--MeerKAT image of the jellyfish galaxy JO147. We show the stellar disk (dashed black contour), the H$\alpha$ emission (orange to black colormap), the radio continuum emission at 1.4 GHz (blue-scale contours, from signal-to-noise of 3 up to 600, angular resolution $12\times12$ arcsec$^{2}$, noise level of 7 $\mu$Jy beam$^{-1}$), and the polarized emission with a signal-to-noise ratio higher than five (green contours). The green contours' intensity and the magnetic field vectors' length (black lines) are proportional to the polarization fraction. The physical scale is shown in the top-right corner, whereas the blue and red circles in the left corner show, respectively, the radio continuum and H$\alpha$ images resolution.}
    \label{image}
\end{figure}

\begin{figure}
    \centering
    \includegraphics[width=\linewidth]{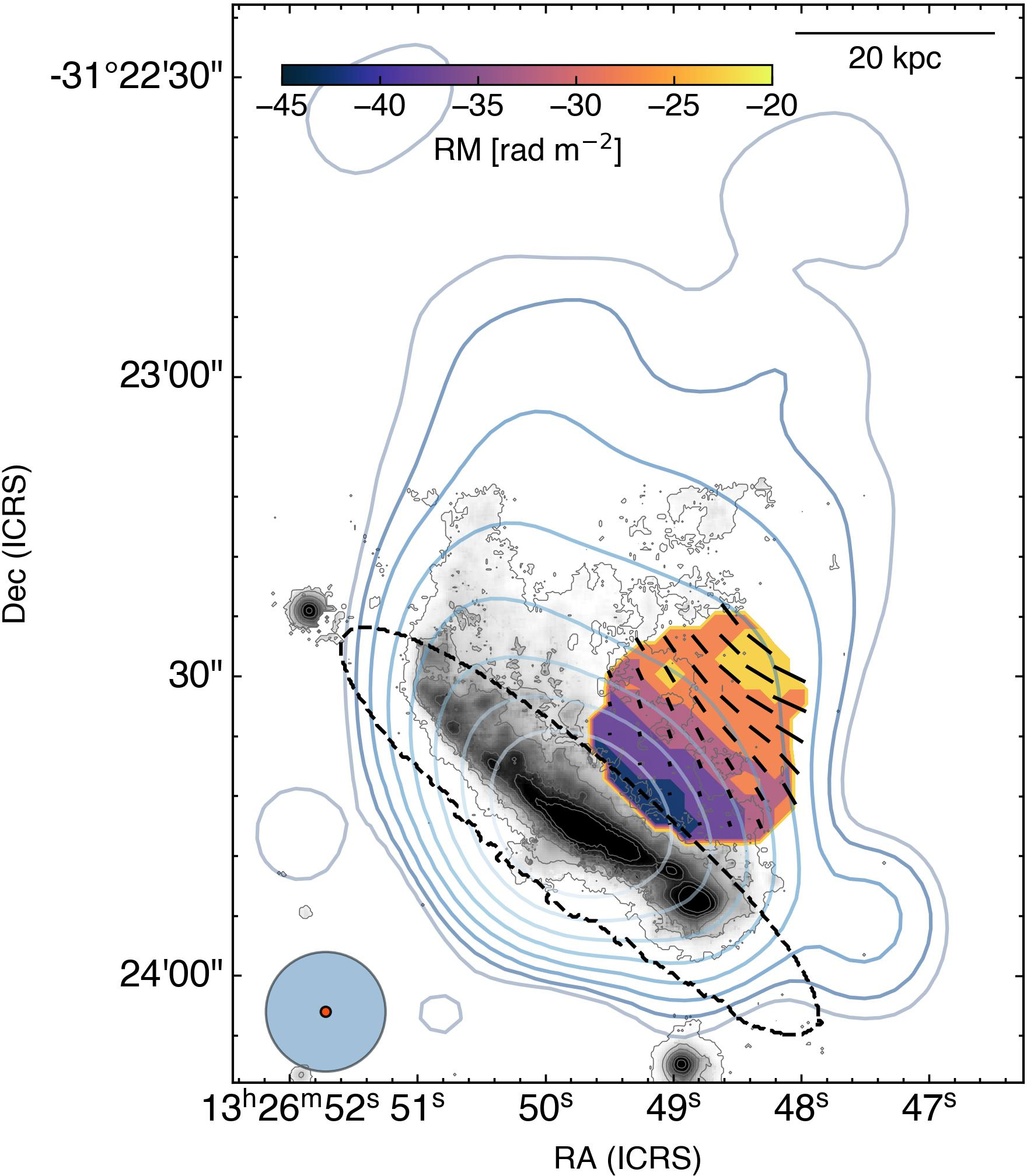}
    \caption{Composite MUSE--MeerKAT image of the jellyfish galaxy JO147. All quantities are identical to those shown in Figure~\ref{image}, except that the color-filled contours now represent the Faraday RM map.}
    \label{RM_map}
\end{figure}

In JO147, MeerKAT detects a 60 kpc long radio continuum tail at 1.4 GHz (Figure \ref{image}), which is consistent with previous findings \citep[][]{2024MNRAS.533.1394M}, and, for the first time, reveals the presence of polarized synchrotron emission in the tail of this galaxy. The polarized emission in JO147 is present only in the western side of the tail and it only marginally overlaps with the stripped warm ISM, traced by the H$\alpha$ emission \citep[][]{Poggianti_2019}. The total polarized flux density is 0.13 mJy, with an average polarized emission fraction of $10\%$. We further observe that the polarization fraction is higher in the region where it does not overlap with the H$\alpha$ emission. The polarized angle vector analysis shows that the magnetic field is mainly oriented parallel to the stellar disk. The polarized signal associated with JO147 is characterized by a Faraday rotation measure (RM) pattern (Figure~\ref{RM_map}) ranging between \mbox{-45} and \mbox{-20} rad m$^{-2}$, with a median value of -30 rad m$^{-2}$, and a tentative gradient along the stripping tail. The Galactic foreground in this region is contributing with only $\sim$4 rad m$^{-2}$ according to \citet{huts}, implying that the signal is either generated by the ICM in front of the galaxy or source-intrinsic. 

\subsection{Differences in magnetic field configurations between JO206 and JO147}
The newly detected polarized emission in JO147 differs from that previously studied in JO206 \citep{Mller2021}, specifically in terms of extension, magnetic field geometry, and Faraday RM signal. We observe that the polarized emission in JO147 is limited to the tail, whereas in JO206 it extends from the stellar disk down to the stripped tail. Common to both galaxies is an increasing polarization fraction from the disk to the tail. Concerning the magnetic field direction, we observe a striking difference where the field appears to be parallel to the stellar disk in JO147 and perpendicular to it in JO206. Furthermore, the two galaxies show a difference in RM. In JO206, \cite{Mller2021} reported, after correcting for the Galactic foreground ($\sim$5.8 rad m$^{-2}$), RM values between -50 and 50 rad m$^{-2}$, which are increasing up to 200 rad m$^{-2}$ towards the tail.

The difference in RM may indicate that the two galaxies resides in different regions of their respective clusters, with JO147 being more peripherical than JO206. This result is in line with the projected clustercentric distance of the two galaxies, which are 0.28 and 0.45$~R_{200}$ for JO206 and JO147 respectively \citep[][]{Gullieuszik2020}. Additionally, the two galaxies show different inclinations with the ICM wind, with JO147 being mostly face-on stripping and JO206 presenting an edge-on configuration. Preliminary numerical simulations of realistic ICM-winds derived from galaxy orbits in cosmological zoom-in cluster simulation \citep[][]{Dusch2025} suggest that the fraction and orientation of the polarized emission in the tail can change with time during the galaxy infall in the cluster, becoming more and more aligned with the stripped material as the galaxy plunge into the cluster. These preliminary results suggests the existence of an evolutionary sequence, which can explain the critical difference between the magnetic field configurations in the two galaxies. Additional work in this direction is hence needed to fully explore how disc inclination, the presence of a galactic wind, gaseous environment and orbit phase affect the magnetic field orientation of the ordered component along the galaxy and its tail. In this work we present a potential framework to interpret both the differences in magnetic field configuration and extraplanar star formation between JO147 and JO206.

\section{Supersonic- vs. ram pressure-driven star formation}
 \subsection{The role of supersonic motions}
 \label{turbo}
 \begin{figure}[t!]
    \centering
    \includegraphics[width=\linewidth]{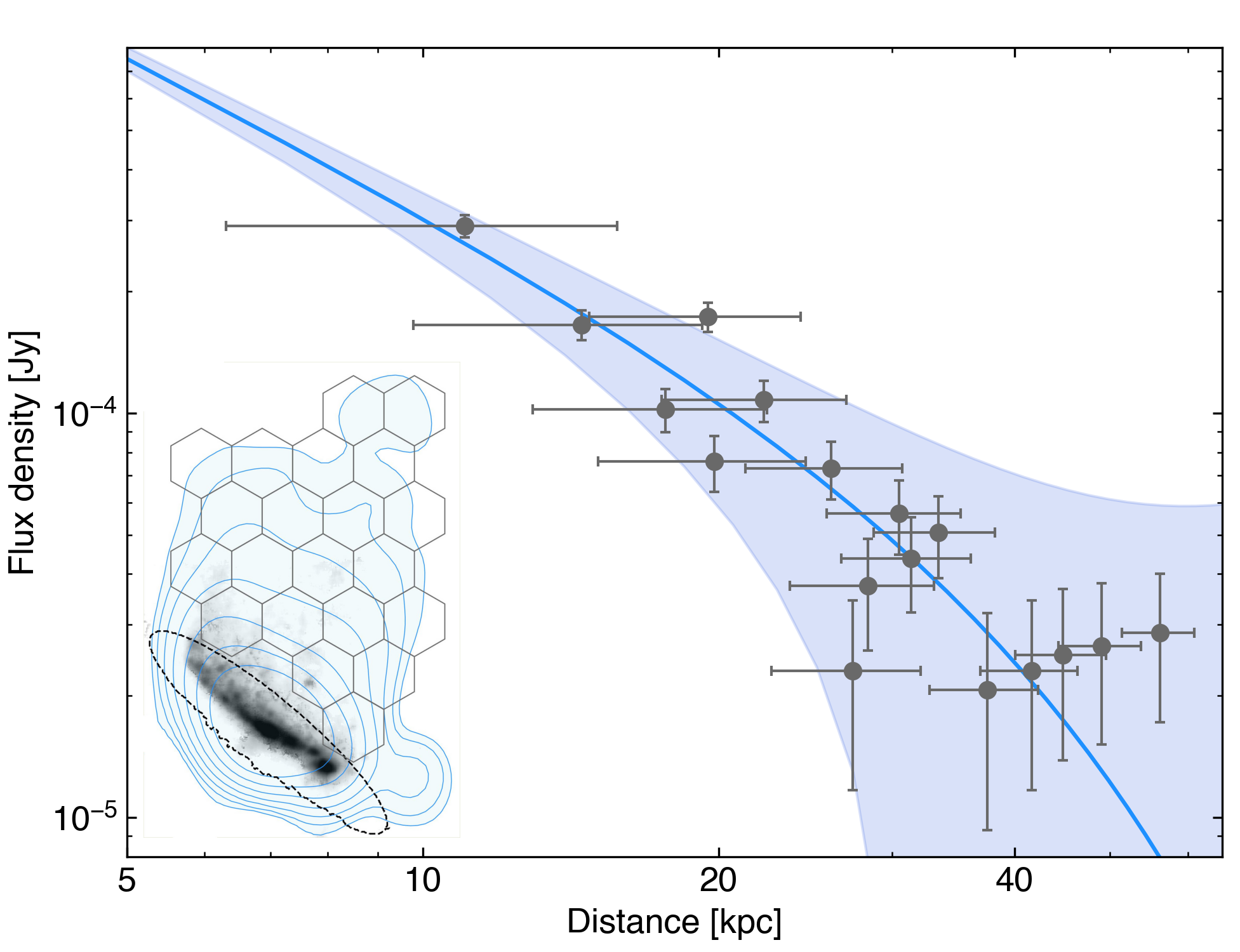}
    \caption{Radio flux density vs.\ distance from the stellar disk edge. In the left corner, we show the sampling grid overlayed on the radio continuum emission shown in Figure \ref{image}. The best-fitting profile is shown by the blue line. The blue-shaded region indicates the $1\sigma$ uncertainties on the fit.}
    \label{image2}
\end{figure}
As outlined in Section \ref{intro}, the extraplanar magnetic field can be composed of two components, an internal one deriving from the stripped ISM magnetic field, and an external one provided by the ICM draping. Ther resulting polarized emission should hence depend on the balance between the intensity of these two components. In the following we present a physical framework to link the observed differences between JO147 and JO206 to a potential different balance between stripped and draped magnetic field intensities. To begin with,  the paucity of coherent polarized emission in JO147, especially in the disk, suggests that the galaxy's magnetic field is mixed with the thermal plasma, which depolarizes most of the nonthermal emission via Faraday rotation. So we detect polarized emission mostly in those regions where the warm ISM, and the corresponding H$\alpha$ emission, are less present. On the contrary, in JO206 the polarized emission was detected from the stellar disk up to the stripped tail. Furthermore, the polarization angle suggests that the magnetic field is mostly aligned with the disk in JO147, instead of the stripped tail. We argue that this is because we are either observing (1) the ISM magnetic field being stripped from the stellar disk, (2) a different upstream magnetic field orientation that is illuminated by cosmic ray electrons in the magnetic drape \citep[][]{Pfrommer2010}, in which case the opposite polarization vectors orientation in JO206 would probe a different upstream orientation of the ICM magnetic field, or (3) an early phase of the magnetic drape formation, with a recent passing of the cluster accretion shock.

JO147 has a projected velocity along the line-of-sight of $664$ km s$^{-1}$ \citep[][]{Gullieuszik2020}, which is larger than the typical ICM Alfv\'en speed, and so it would be in the condition to form a magnetic drape. Hence, we argue that the lack of abundant coherent continuum emission from it can be due to two factors, either a scarcity of cosmic ray electrons coming from the stellar disk to light up the magnetic drape, or a low drape magnetic field energy density, hence synchrotron emissivity. Both JO147 and JO206 are currently forming stars in their disk \citep[][]{Vulcani_2018,Gullieuszik2020} and show extended radio continuum tails, which indicates that in both systems there are cosmic rays available to illuminate the drape, hence challenging the first possibility. On the other hand, the drape magnetic energy density, $\epsilon_{B}$, depends on the galaxy velocity \citep[][]{Dursi2008} and it can be expressed as $\epsilon_{B}=\alpha\gamma\mathcal{M}^2P_{\mathrm{ICM}}$, where $\alpha\simeq2$ is the geometrical factor \citep[][]{Dursi2008}, $\mathcal{M}=V/c_s$ is the Mach number, where $V$ is the galaxy velocity and $c_s$ is the local sound speed, $\gamma$ is the ICM adiabatic index and $P_{\text{ICM}}$ is the ICM thermal pressure. If the galaxy has reached a sufficient velocity to become supersonic, i.e. $\mathcal{M}=V/c_s>1$, then the shock compression can further amplify the field strength \citep[][]{Sparre2020}. Therefore, we argue that the key factor that differentiates the two galaxies and leads to different properties in their tail is that JO147 experienced a weaker shock compression than JO206, resulting from a low-Mach motion in the ICM. As a direct consequence, JO147 has not formed a strong magnetic drape yet which, in turn, would have resulted an extraplanar magnetic field dominated by the stripped ISM, hence limited polarized emission, and fast evaporation of the stripped material, due to the lack of the protective magnetic drape. Supersonic motions can also have another implication. The resulting counter shock crossing the galaxy can impose a compression on the ISM, which depends on the shock Mach number. The compression increases the ISM density, which results in more efficient cooling. In case of significant ISM compression, the dense stripped clouds will efficiently cool down, thus becoming resilient to thermal evaporation \citep[][]{Gronke2018}. In this framework, a weak shock compression in JO147 would have resulted in both stripped ISM clouds being less resilient to the external heating, due to the lower density, and a weak magnetic drape that could not shield the stripped gas from the ICM. 

To test this hypothesis, it is necessary to constrain the JO147 3D galaxy velocity with respect to the ICM and the corresponding Mach number. The velocity component along the line-of-sight can be derived from the spectroscopic optical observation of the galaxy, whereas the perpendicular component, which is consistent with stripped ISM velocity along the plane of the sky \citep[][]{Roberts_2024}, can be constrained from the curvature of the radio continuum gradient along the stripped tail by following the approach described in \cite[][]{Ignesti2023}. The radio emission is sampled using the \texttt{PT-REX} code\footnote{\url{https://github.com/AIgnesti/PT-REX}} \citep[][]{Ignesti2022b} with a hexagonal grid, where each bin is as large as the resolution of the radio continuum image, $12\times12$ arcsec$^{2}$. The grid covers the radio emission outside of the stellar disk and with a signal-to-noise ratio higher than 3. To compose the radio flux density profile, for each bin, we measure the radio continuum flux density and the projected distance from the stellar disk, which we measure as the shortest path from the center of the cell to the stellar disk mask edge. The profile is then fitted with the semi-empirical model derived under the assumption that 1) the relativistic electrons are accelerated only in the stellar disk which they leave on a time-scale significantly shorter than the energy loss timescale, 2) the energy losses outside of the stellar disk are dominated by synchrotron and Inverse Compton radiation, 3) their bulk motion is described by a uniform velocity along the stripping direction, and 4) they move in a uniform magnetic field \citep[][]{Ignesti2023}. The model is fitted to the observed profile by using the least squares method to derive the best-fitting velocity and amplitude in flux density units, as well as the associate uncertainties. 

The radio continuum flux density profile along the tail (Figure \ref{image2}) shows a monotonic decline with distance. By adopting a typical range for magnetic field intensity $B=2-5$ $\mu$G along the stripped tails \citep[][]{Ignesti2022,Roberts_2024}, the best-fitting profile returns an average velocity along the plane of the sky in the range $V_{\perp}=762-1089$ km s$^{-1}$, which, added in quadrature with the velocity component along the line of sight, $V_{\parallel}=664$ km s$^{-1}$ \citep[][]{Gullieuszik2020}, derived from the MUSE spectroscopical observations, results in a total velocity of $V=\sqrt{V_{\perp}^2+V_{\parallel}^2}=1013-1278$ km s$^{-1}$. Previous studies \citep[][]{Bardelli1996} of the ICM thermal properties at the galaxy clustercentric distance ($13.8$ arcmin, corresponding to ${\sim}876$ kpc at the cluster redshift) permit us to constrain the local sound speed at $c_s\simeq800$ km s$^{-1}$, resulting in a Mach number in the range $\mathcal{M}=1.3-1.6$, which entails a weak compression in accordance with our hypothesis. 

We note that, as discussed in \citet[][]{Ignesti2023}, a crucial caveat of this method is the non-linear magnetic field intensity-velocity degeneracy due to the fact that a strong/weak magnetic field entails short/long cooling time-scales, hence a high/low velocity required for the radio plasma to extend over the observed radio tail length. For reference, JO147 would achieve $\mathcal{M}>2$ only for $B>7$ $\mu$G, which would be larger than typically inferred for this class of galaxies \citep[][]{Mller2021,Ignesti2022,Roberts_2024}. We further observe that a direct comparison with JO206 is not possible because \citet[][]{Mller2021} could only derive a lower limit $\mathcal{M}>1.3$, under the assumption that the velocity component along the line of sight was equal to the perpendicular component, albeit the galaxy morphology suggests that the perpendicular component is significantly larger than the parallel one \citep[][]{Jaffe_2018}. Moreover, the radio emission gradient analysis can not be performed on JO206 because, due to the high extraplanar star formation, the assumption of relativistic electrons originating only in the disk is not respected.

\subsection{Testing the general implications}
\label{test_MC}
\begin{figure}
    \centering
    \includegraphics[width=.95\linewidth]{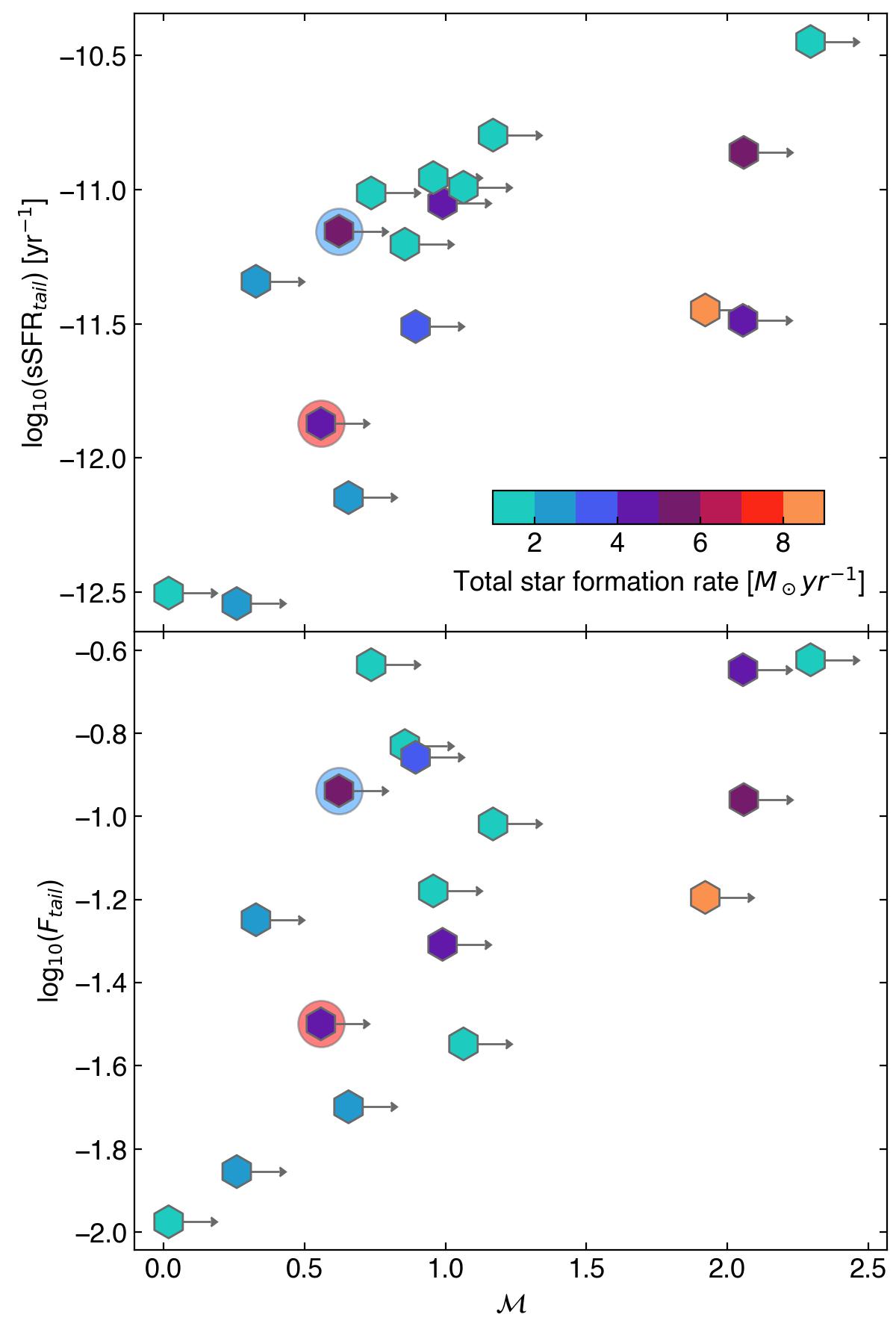}
    \caption{Extraplanar star formation efficiency vs.\ ``projected'' Mach number.  Extraplanar specific star formation rate, log$_{10}(\mathrm{sSFR}_{\text{tail}})$ (top) and star formation rate fraction, log$_{10}(F_{\text{tail}})$ (bottom), vs.\ projected Mach number lower limits, $\mathcal{M}$. Each galaxy is color-coded for its corresponding total star formation rate \citep[][]{Gullieuszik2020}. JO147 and JO206 are marked with red and blue circles, respectively. }
    \label{fig_proj}
\end{figure}

In our proposed framework, jellyfish galaxies star-forming tails result from the shock compression they experience during their orbits. It would then follow that galaxies with a higher Mach number should exhibit a relatively higher amount of star formation in their tails. To test this prediction, we examine a sample of jellyfish galaxies from the GASP sample \citep[Gas Stripping Phenomena in Galaxies with MUSE, ][]{Poggianti2017,Poggianti2025} to compare their capability in forming stars outside of their stellar disk with their Mach number. In principle, an accurate measure of a cluster galaxy Mach number would require high-angular resolution X-ray observations to determine the the morphology and the spectral properties of the bow-shock induced in the ICM \citep[e.g.,][]{Wez_2011,Poggianti2019}, in which temperature and density jump between upstream and downstream are directly related to the shock Mach number. However, this measurement is often contaminated by projection effects, which can smooth out the thin temperature and density jump induced by the galaxies, or the generally low surface brightness, which makes difficult to determine the spectral properties of the up- and downstream regions. Therefore, in this work we propose to analyze the ``statistical'' Mach number. First, as reference, we compute the ``projected'' Mach number, which we calculate as the line-of-sight velocity component divided by the average ICM sound speed, which is based on the average ICM temperature. The star formation rate of GASP galaxies, as well as the amount of it taking place outside of the stellar disks, have been previously calculated from MUSE observations \citep[][]{Gullieuszik2020}. Their relative amount of extraplanar star formation is estimated from both the fraction of star formation in the tail, $F_{\text{tail}}$, which comes from the ratio between the extraplanar and the total star formation, and the extraplanar specific star formation rate, $\mathrm{sSFR}_{\text{tail}}$, derived by dividing the extraplanar star formation rates by the corresponding total stellar mass, $M_*$, which correlates with the total star formation \citep[][]{Vulcani_2018}. The final sample is composed of 17 spiral galaxies in a state of strong or extreme stripping \citep[Morphological type $T>-1$ and Jtype=1,2, see][]{Vulcani_2015, Poggianti2025}, with $F_{\text{tail}}>0.01$, to ensure a minimum amount of extraplanar star formation, and a projected cluster-centric distance lower than $R_{500}$, the physical radius within which the average cluster density exceeds 500 times the critical universe density at that redshift.

From this comparison, shown in Figure \ref{fig_proj}, a positive trend between the ``projected'' Mach number and the extraplanar star formation efficiency emerges, which is in line with our hypothesis. However, the projected Mach number is a lower limit to its real value. In order to further test the trends shown in Figure \ref{fig_proj}, we use a statistical method to infer the 3D velocity and cluster-centric distance of GASP jellyfishes by matching their location in the phase-space with those produced by a realistic family of orbits integrated within the cluster gravitational potential.

For each galaxy, we simulate the possible 3D orbits within the hosting cluster to derive the distributions of 3D velocity and position to associate with their projected counterparts. The hosting cluster gravitational potential is modeled as an NFW profile \citep[][]{Navarro1997}, where the mass and concentration parameters, namely $M_{200}$ and $c$, have been previously derived in \cite[][]{Biviano2017}. The integration of each orbit has been carried out with the \texttt{gala} Python package\citep[][]{gala}. Every orbit starts from a random point on a sphere with radius equal to $R_{200}$ from the cluster center with the infall velocities randomly sampled from the distributions of tangential and radial velocities, rescaled for the hosting cluster velocity dispersion, presented in \cite{Smith2022b}. These distributions are derived from the orbits of dark matter halos on their first infall into a sample of nearly 740 groups and clusters in cosmological simulations. For each of the 42353 infalling galaxies, the tangential and radial component of their velocity was measured at the moment of crossing $R_{200}$ of the group or cluster. To replicate the typical radial orbits of jellyfish galaxies \citep[][]{Wetzel_2010,Rhee_2017, Biviano_2024}, only orbits with a ratio between the initial radial and tangential components larger than 1.4 are selected. Relaxing this assumption result in generally lower Mach number estimates but it does not change the trends. Observational evidence suggests that optically-selected jellyfish galaxies preferentially form their tail during their first infall \citep[][]{Smith2022, Salinas_2024}. Accordingly, each orbit is simulated until it reaches the pericenter, and then it is projected into phase-space coordinates by considering the distance along the x-axis as the projected cluster-centric distance and the velocity along the y-axis as the line-of-sight velocity. When a projected orbit crosses the observed phase-space coordinates, within an interval of 10$\%$ of their observed values, the corresponding values of 3D velocity and distance are stored. 

\begin{figure}
    \centering
    \includegraphics[width=\linewidth]{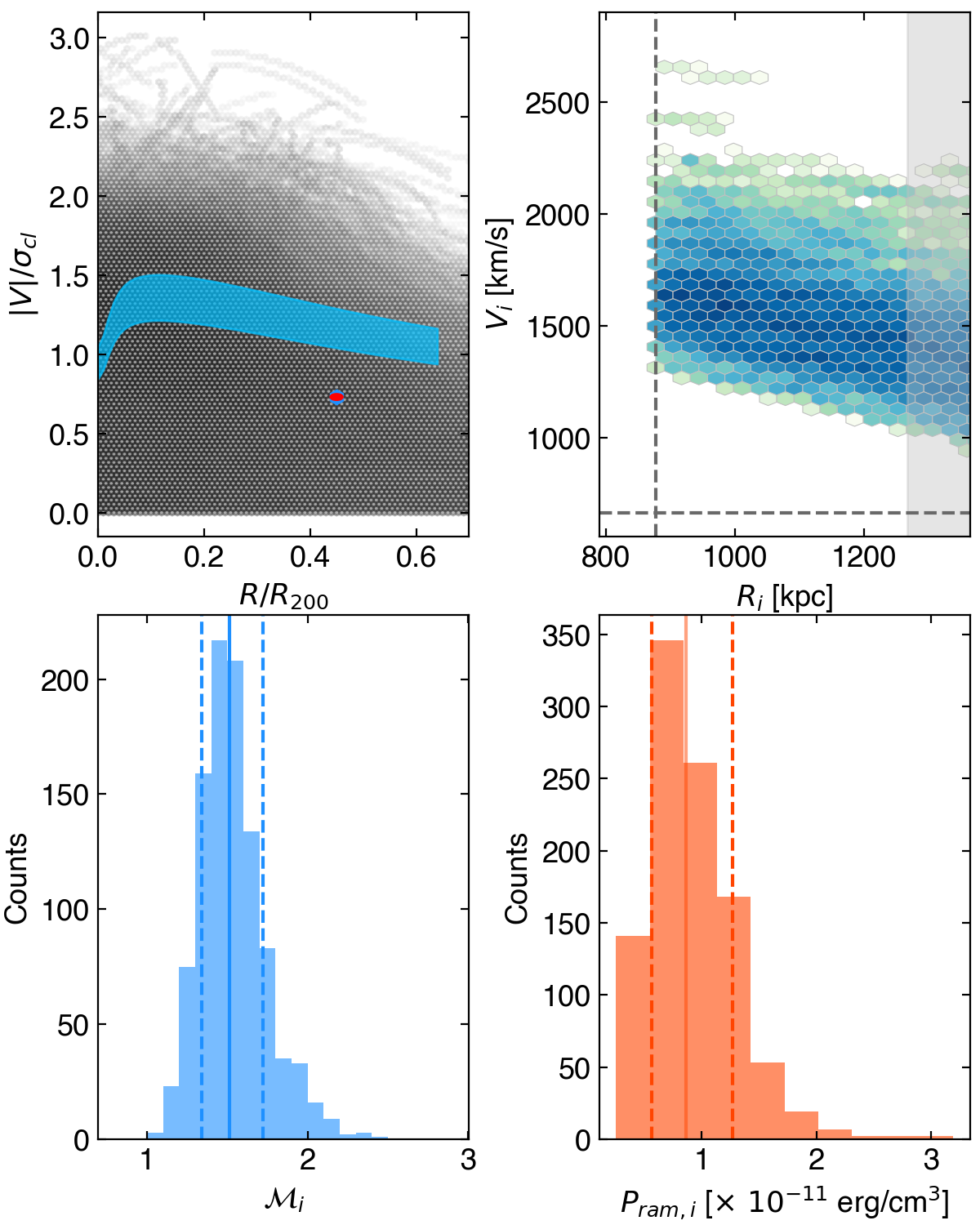}
    \caption{Outcomes of Monte Carlo analysis for JO147. Top: Simulated orbits projected on the phase-space plane, with the red dot showing the projected phase-space coordinates for JO147 and the blue line indicating the corresponding ICM sound speed profile (left) and 2D Distribution of the $V_i$-$R_i$ pairs associated with the projected JO147 coordinates, the dashed lines indicate JO147 projected coordinates, and the shaded area covers the rejected $V_i$-$R_i$ solutions (right); Bottom: $\mathcal{M}_i$ (left) and $P_{ram,i}$ (right) distributions, the vertical lines indicate the median and the 16$^{\text{th}}$ and 84$^{\text{th}}$ percentiles. }
    \label{MC_outcome}
\end{figure}

To compute the Mach number corresponding to a given pair of 3D velocity-distance coordinates, it is necessary to know the local ICM sound speed, which, under the assumption of monoatomic thermal gas, depends only on the ICM temperature as $c_s\simeq515\sqrt{(kT/1\text{ keV})}$ km s$^{-1}$. The ICM temperature profile is modeled using the averaged analytical profile presented in \cite[][]{Vikhlinin2006}. The average ICM temperature of each cluster is estimated as the median of the values reported in previous X-ray studies \citep[][]{David_1993,Cavagnolo2009,Sanderson2006,Poggianti2019,Mller2021,Bartolini2022,Bulbul2024}. For each orbit, the temperature profile is assumed to vary within a $15\%$ scatter resulting from the analytical profile uncertainty. The analytical temperature profile is valid only within $R_{500}$, which represents an additional condition to select the 3D velocity-distance pairs. The corresponding 3D Mach number for each orbit is finally computed as $\mathcal{M}_i=V_i/c_s(R_i)$, where the $i$ subscript indicates the iteration. We repeat the simulation until we collect a distribution of 1000 Mach values per galaxy or when we run $10^{5}$ orbits. The final Mach number estimates, and their uncertainties, are computed from the median and the 16$^{\text{th}}$ and 84$^{\text{th}}$ percentiles of the distribution. The ram pressure $P_{\text{ram}}$ is computed by modeling the ICM density profile following \citet[][]{Pratt_2022}, including the scatter on the best-fitting parameters. Similarly to the case of the Mach number, we derive a distribution of $P_{\text{ram}}$ composed of each 3D velocity-distance combination consistent with their projected counterparts, as $P_{\text{ram},i}=\rho(R_i)V_i^2$. Then the final estimates, and their uncertainties, are computed from the median and the 16$^{\text{th}}$ and 84$^{\text{th}}$ percentiles of the distributions. In Table \ref{table}, we report the input parameter for each galaxy, the resulting estimates of $\mathcal{M}_\text{3D}$ and $P_{\text{ram}}$ and the percentage of success, $S_p$, computed as the ratio between the number of orbits crossing the phase-space coordinates and the total number of simulated orbits. As an example, we show the Monte Carlo analysis steps for the galaxy JO147 in Figure \ref{MC_outcome}. Specifically, we show the projected simulated orbits in comparison to the galaxy phase-space coordinates (top-left panel), the the distributions of 3D velocities ($V_i$) and cluster-centric distances ($R_i$) associated with the projected coordinates (top-right panel), and the resulting  $\mathcal{M}_i$ and $P_{\text{ram},i}$ distributions (bottom panels).

\begin{table*}[]
\setlength{\tabcolsep}{4pt}

    \centering
    {\renewcommand{\arraystretch}{1.5}%

    \begin{tabular}{cccccccccc}
\toprule
\midrule
Galaxy&$R$&$V$&SFR&$F_{\text{tail}}$&$\mathrm{sSFR}_{\text{tail}}$&$kT_X$&$\mathcal{M}_\text{3D}$&$P_{\text{ram}}$&$S_p$\\
&[$R_{200}$]&[$\sigma_{\text{cl}}$]&[$\text{M}_\odot$/yr]&[log$_{10}$]&[log$_{10}$]&[keV]&&[$\times10^{-11}$ erg cm$^{-3}$]&[$\%$]\\
\midrule
JO49&0.41&0.03&1.42&-1.98&-12.5&2.6&1.1$_{-0.1}^{+0.2}$&0.17$_{-0.06}^{+0.08}$&1.18\\
JO60&0.49&1.77&4.55&-1.31&-11.05&4.04&1.3$_{-0.1}^{+0.2}$&0.34$_{-0.07}^{+0.11}$&0.99$^*$\\
JO85&0.18&1.69&5.87&-0.96&-10.86&3.4&2.6$_{-0.2}^{+0.3}$&6.49$_{-3.88}^{+6.55}$&12.44\\
JO95&0.25&1.15&0.39&-1.18&-10.96&3.81&2.0$_{-0.2}^{+0.2}$&1.84$_{-1.14}^{+2.64}$&11.04\\
JO113&0.58&1.28&1.76&-1.55&-10.99&5.03&1.7$_{-0.1}^{+0.2}$&0.75$_{-0.14}^{+0.24}$&1.31\\
JO135&0.16&0.4&2.0&-1.85&-12.54&3.96&1.4$_{-0.1}^{+0.1}$&1.62$_{-0.96}^{+1.55}$&4.55\\
JO147&0.45&0.73&4.55&-1.5&-11.87&5.44&1.5$_{-0.2}^{+0.2}$&0.86$_{-0.32}^{+0.38}$&4.7\\
JO162&0.41&1.34&0.44&-1.02&-10.8&3.2&1.8$_{-0.2}^{+0.2}$&0.77$_{-0.28}^{+0.42}$&5.12\\
JO171&0.62&0.87&1.71&-0.64&-11.01&5.68&1.7$_{-0.2}^{+0.3}$&0.84$_{-0.16}^{+0.28}$&1.47\\
JO175&0.28&0.29&2.55&-1.25&-11.34&1.7&2.4$_{-0.2}^{+0.2}$&1.28$_{-0.68}^{+0.77}$&3.38\\
JO194&0.17&2.62&8.44&-1.19&-11.45&3.96&2.1$_{-0.2}^{+0.2}$&3.93$_{-2.18}^{+3.37}$&0.05$^*$\\
JO200&0.46&0.99&2.35&-1.7&-12.15&6.44&1.4$_{-0.1}^{+0.2}$&0.85$_{-0.28}^{+0.39}$&4.79\\
JO204&0.08&1.18&1.72&-0.83&-11.2&2.9&1.5$_{-0.1}^{+0.1}$&1.3$_{-0.85}^{+1.97}$&10.81\\
JO206&0.28&1.09&5.54&-0.94&-11.15&3.9&1.1$_{-0.1}^{+0.1}$&0.42$_{-0.24}^{+0.35}$&9.38\\
JW39&0.33&1.25&3.61&-0.86&-11.51&3.18&1.7$_{-0.2}^{+0.2}$&0.7$_{-0.37}^{+0.72}$&8.91\\
JW56&0.16&2.33&0.17&-0.62&-10.45&3.36&2.5$_{-0.1}^{+0.2}$&8.35$_{-3.68}^{+5.34}$&7.47\\
JW100&0.06&2.95&4.26&-0.65&-11.49&3.29&2.3$_{-0.0}^{+0.2}$&2.4$_{-0.74}^{+5.38}$&0.01$^*$\\
\bottomrule
    \end{tabular}}
    \quad
    \caption{Galaxy sample parameters. From left to right: GASP name; Normalized projected clustercentric distance; Normalized line-of-sight velocity; Total star formation rate; Extraplanar star formation rate fraction; Extraplanar specific star formation rate; Average hosting cluster X-ray temperature; Resulting Mach number; Resulting Ram pressure; Monte-Carlo success ratio. Note: galaxies for which the desired number of Mach number estimates were not achieved within $10^5$ orbits are denoted with an asterisk ($^*$).}
    \label{table}
\end{table*}

\begin{figure}
    \centering
    \includegraphics[width=\linewidth]{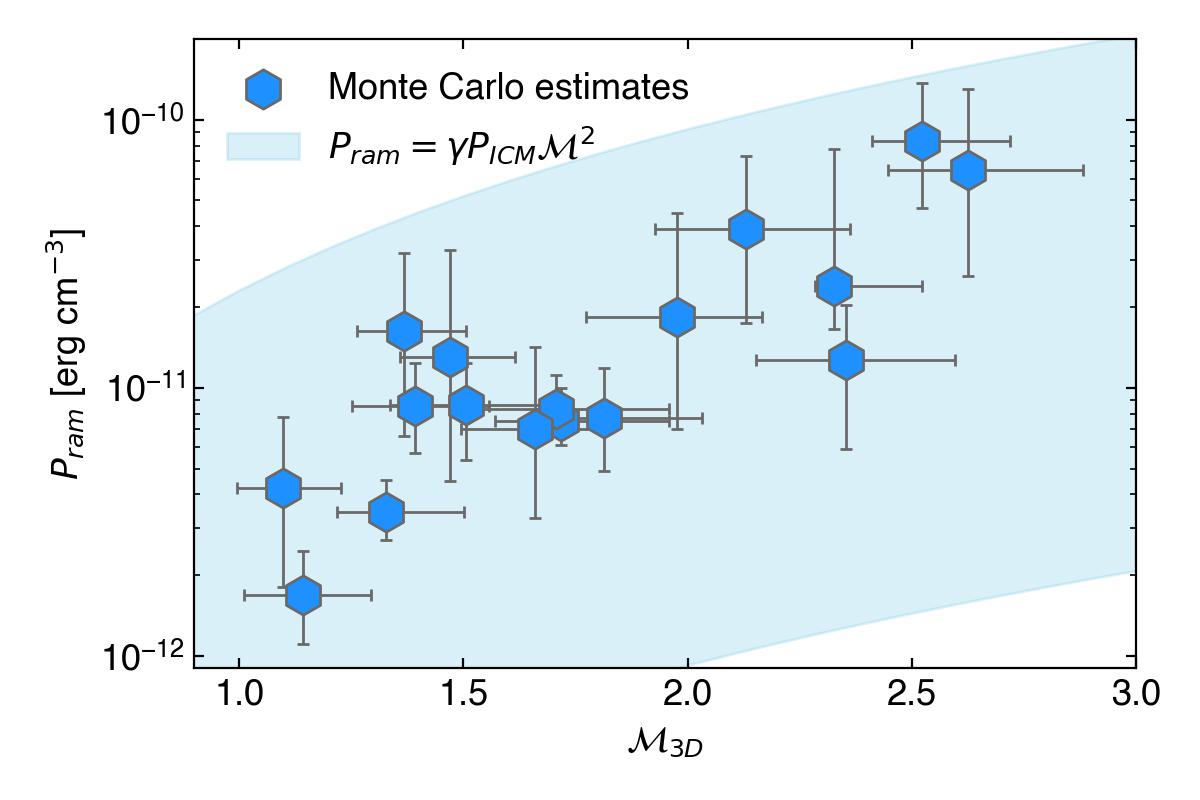}
    \caption{Values of $\mathcal{M}_{\text{3D}}$ and $P_{\text{ram}}$ inferred with the Monte Carlo analyisis, compared with the expected trend for typical ICM thermal properties (Equation \ref{test_eq}).}
    \label{test_mach}
\end{figure}

For GASP jellyfish galaxies, we infer $1<\mathcal{M}_{\text{3D}}<2.6$, which would result in a maximum compression factor of $\sim1.9$, and $-12<\text{log}_{10}(P_{\text{ram}}/\text{erg cm}^{-3})<-10$, which are consistent with previous estimates \citep[][]{Wez_2011,Yun_2019,Ignesti2023}. As an additional sanity check, in Figure \ref{test_mach} we show that the values of Mach number and ram pressure inferred with the Monte Carlo  analysis for each galaxy follow the expected non-linear relation:
\begin{equation}
P_{\text{ram}}=\rho V^2=\rho~c_s^2 \mathcal{M}^2=\rho\left(\sqrt{\gamma\frac{P_{\text{ICM}}}{\rho}} \right)^2\mathcal{M}^2=\gamma~P_{\text{ICM}}\mathcal{M}^2\text{,} 
\label{test_eq}
\end{equation}
where $\gamma=5/3$ is the adiabatic index and $P_{\text{ICM}}$ is the ICM thermal pressure computed for typical ICM particle density between $10^{-4}$ and $10^{-3}$ cm$^{-3}$ and temperature between $10^7$ and $10^8$ K. 
For JO147, the Monte Carlo estimate $\mathcal{M}_{\text{3D}}=1.5\pm0.2$ is consistent with the measurement based on the radio continuum gradient, $\mathcal{M}=1.3-1.6$. Regarding JO206, we infer $\mathcal{M}_{\text{3D}}=1.1\pm0.1$. However, we note that it is one of the cases where $\mathcal{M}_\text{3D}$ may be systematically under-estimated (for a detailed discussion we refer to Section \ref{met3}). In general, with the Monte-Carlo approach we infer $\mathcal{M}_{\text{3D}}$ which are systematically higher than the corresponding projected values, with the difference being larger for those galaxies showing the lowest line-of-sight velocity.

\begin{figure*}
\centering
    \includegraphics[width=0.75\textwidth]{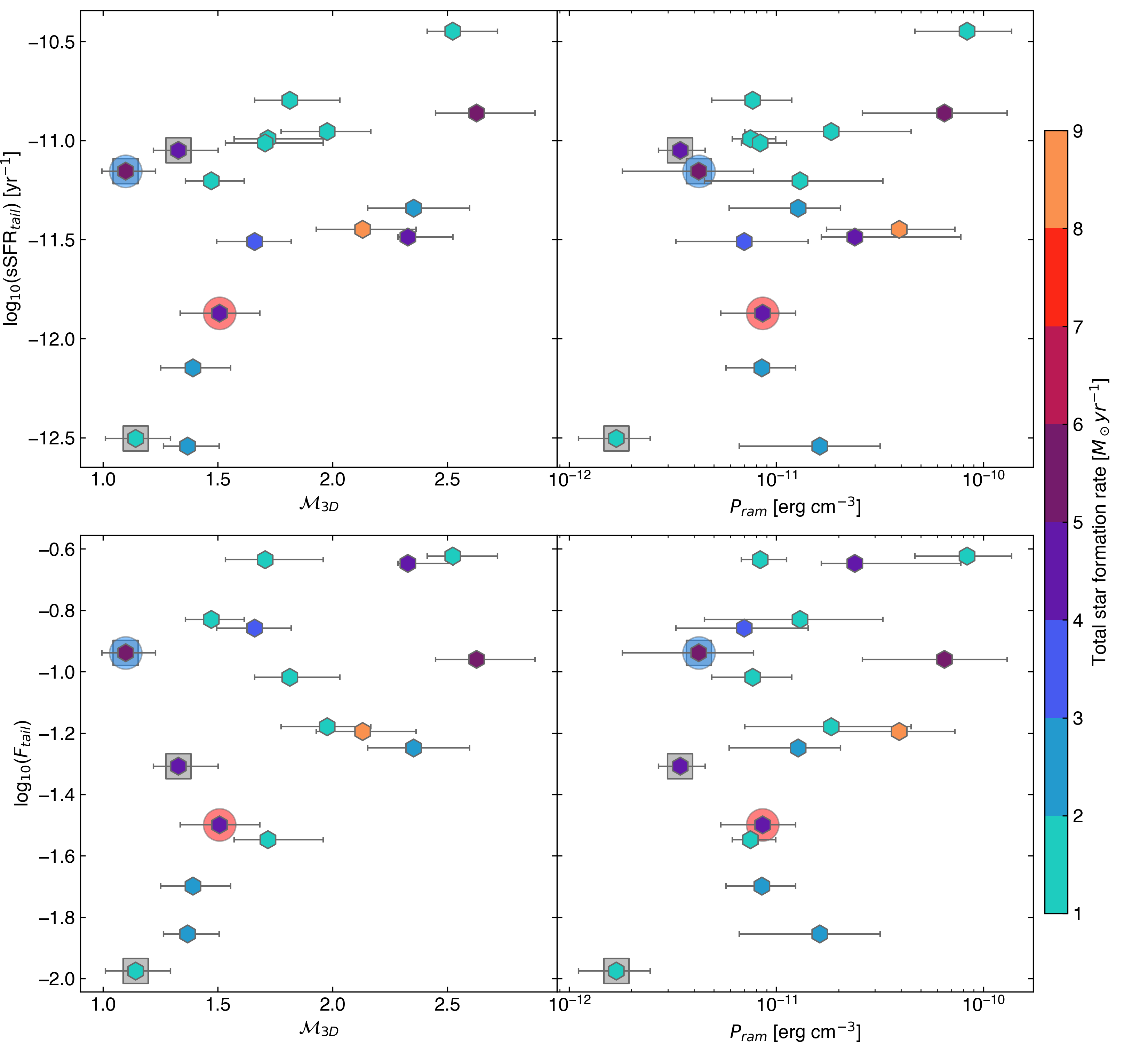}
    \caption{Extraplanar star formation efficiency vs.\ dynamical properties of the GASP sample. Extraplanar specific star formation rate, log$_{10}(\mathrm{sSFR}_{\text{tail}})$ (top) and star formation rate fraction, log$_{10}(F_{\text{tail}})$ (bottom), vs.\ 3D Mach number, $\mathcal{M}_{\text{3D}}$ (left), and ram pressure, $P_{\text{ram}}$ (right, presented on a logarithmic scale for visual clarity). Each point is color-coded for its corresponding total star formation rate \citep[][]{Gullieuszik2020}. JO147 and JO206 are marked with red and blue circles, respectively. Points marked with silver boxes may suffer from systematic $\mathcal{M}_\text{3D}$ under-estimates (see Section \ref{met3}). }
    \label{phase-space}
\end{figure*}

\begin{figure*}
\centering
    \includegraphics[width=.45\linewidth]{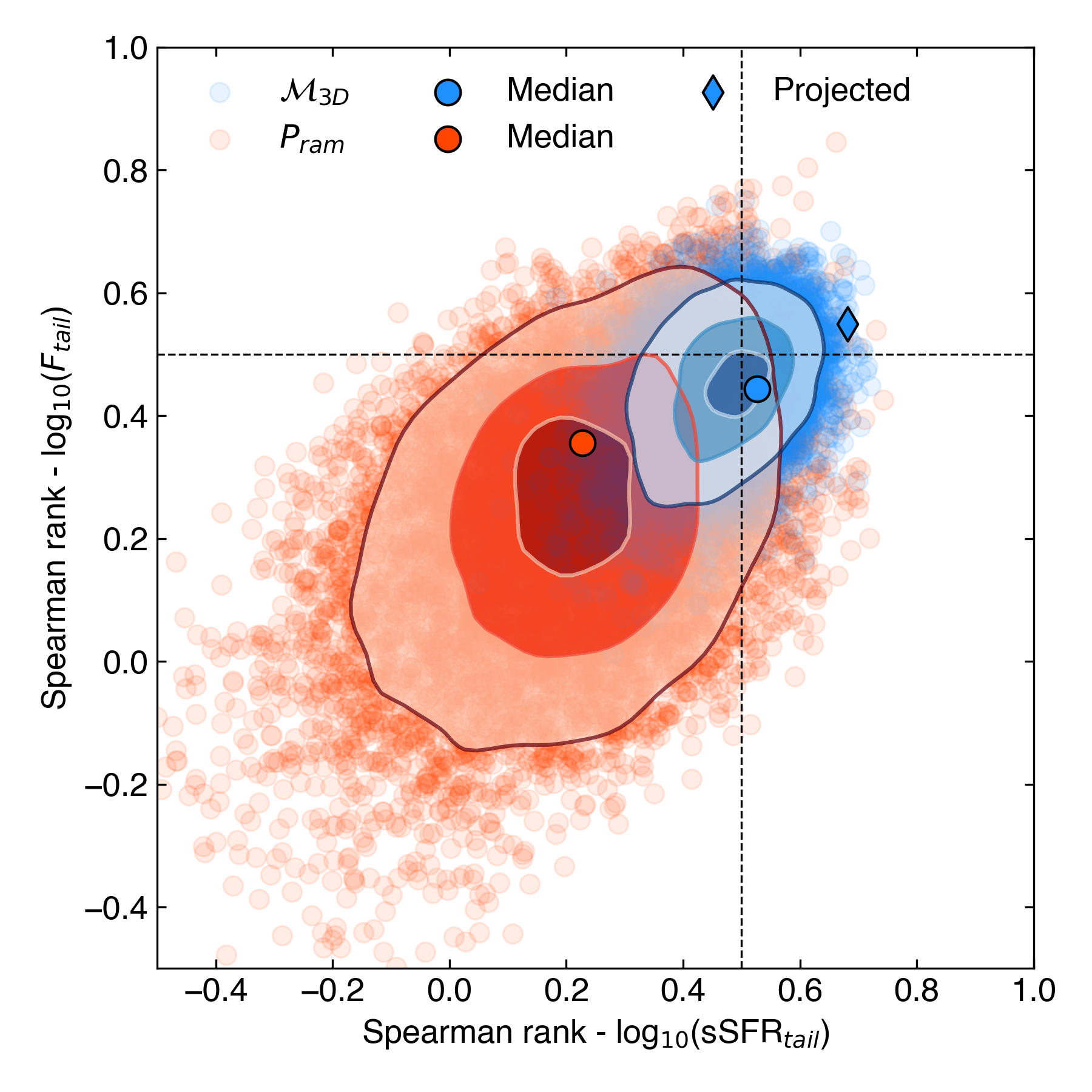}
    \includegraphics[width=.45\linewidth]{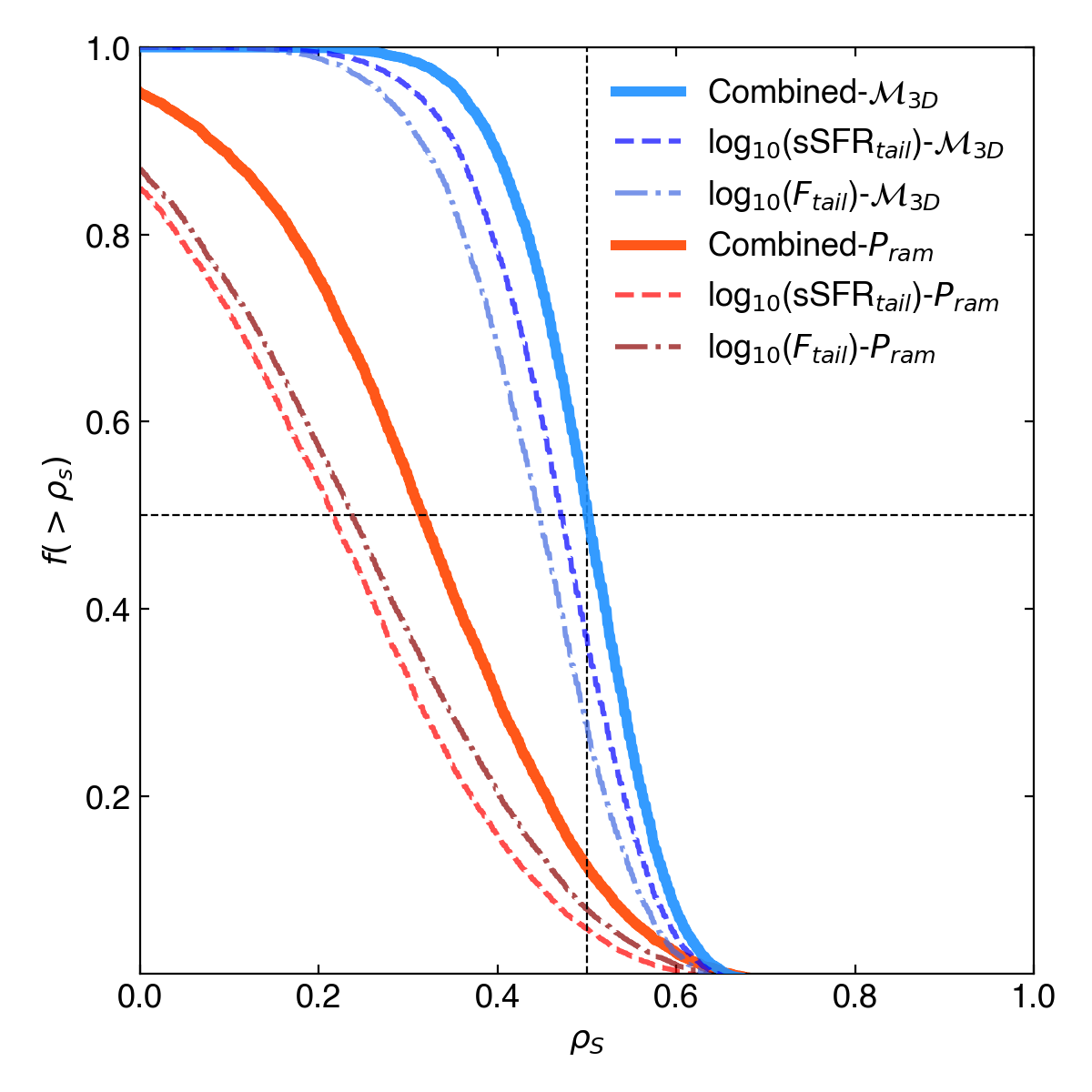}
    \caption{Spearman rank distributions. Left: Spearman rank distributions derived for log$_{10}(\mathrm{sSFR}_{\text{tail}})$ and log$_{10}(F_{\text{tail}})$ when compared with $\mathcal{M}_{\text{3D}}$ (blue) and $P_{\text{ram}}$ (red). Each point shows the ranks derived for different realizations of the $\mathcal{M}_{\text{3D}}$ and $P_{\text{ram}}$ series. The contours indicate the 2D probability density levels of 0.16, 0.5 and 0.84. The black-dashed lines indicate the $\rho_s=+0.5$, corresponding to an existing, positive correlation between the two quantities. The black-outlined points indicate the Spearman ranks measured from the median values reported in Figure \ref{phase-space}. For reference, the blue diamond indicates the correlation rank from the ``projected'' $\mathcal{M}$ (Figure \ref{fig_proj}); Right: fraction of realizations with a Spearman rank higher than $\rho_s$ for $\mathcal{M}_{\text{3D}}$ (blue) and $P_{\text{ram}}$ (red). The solid lines show the combined fractions of log$_{10}(\mathrm{sSFR}_{\text{tail}})$ and log$_{10}(F_{\text{tail}})$, and the dashed and dot-dashed lines show the fractions for the individual quantities. The black dashed lines indicate the 0.5 levels on the corresponding axis.} 
    \label{spear}
\end{figure*}

In Figure, \ref{phase-space} $\mathrm{sSFR}_{\text{tail}}$ (top) and $F_{\text{tail}}$ (bottom) are compared with $\mathcal{M}_{\text{3D}}$ (left) and $P_{\text{ram}}$ (right) to determine which factor, shock compression or ram pressure, plays a more significant role in triggering extraplanar star formation. For reference, galaxies are color-coded for their total star formation rate. We further quantify the correlation strength by computing the distribution of Spearman rank $\rho_s$ from 10000 random realization of our dataset, each built by extracting the relevant quantities, $\mathcal{M}_{\text{3D}}$ and $P_{\text{ram}}$, from normal distributions with mean and standard deviation given by each estimate and associated uncertainties. The resulting $\rho_s$ distribution, shown in Figure \ref{spear}, supports a stronger correlation of both $F_{\text{tail}}$ and $\mathrm{sSFR}_{\text{tail}}$ with $\mathcal{M}_{\text{3D}}$ than with $P_{\text{ram}}$: we find $\rho_s>0.5$ for 51$\%$ of the realizations using $\mathcal{M}_{\text{3D}}$ (Figure \ref{spear}, right panel), with an average correlation rank $\rho_s=0.5\pm0.1$, and only for 13$\%$ of the realizations using $P_{\text{ram}}$, which results instead in an average $\rho_s=0.2\pm0.2$.

\subsection{Caveats}
\label{met3}
The Monte Carlo analysis presented in this work is subject to a number of caveats and limitations to be taken into consideration.
\begin{itemize}
    \item  We observe that the constraint $R_i\leq R_{500}$ sets a systematic upper limit for the $V_i$ and $R_i$ distributions and, correspondingly, it narrows the resulting $\mathcal{M}_i$ distribution.
    \item For reference, we also compute the Mach number in the case of an isothermal ICM, corresponding to a case where the sound speed is uniform in the cluster. This model overestimates the sound speed at large distances, hence resulting in a generally lower Mach number values than the analytical profile case.
    
    \item We note that this approach assumes spherical symmetry, the cluster dynamical equilibrium, and the infalling velocity normalization to the cluster velocity dispersion. These assumptions may not be a solid representation for systems undergoing mergers, for which the actual temperature is higher than the equilibrium one, or for extremely fast infalling galaxies. In these cases, the Mach number can be underestimated. This effect may potentially explain the low values of $\mathcal{M}_\text{3D}$ we infer for some of the high-star-forming galaxies, signed with silver boxes in Figure \ref{phase-space}, which are hosted in clusters for which the observed temperature is higher than the one predicted by the mass-temperature scaling relation \citep[][]{Babyk2023}. The case of JO206 appears to be especially critical, as it would be a fast infaller in a relatively low mass cluster \citep[$M_{200}\simeq2\times10^{14}~M_\odot$,][]{Gullieuszik2020} with, in contrast, a relatively high average ICM temperature of 3.9 keV, that is $\sim1.4$ times higher than expected given its mass. As a result, our model predicts relatively low velocities with respect to the local ICM sound speed, thus resulting in a low Mach number estimate. Correspondingly, in the so-called cool-core systems, the central temperature can be lower than the one predicted by the analytical model, resulting in an overestimate of the Mach numbers for the orbits crossing the cluster center.
    \item With this approach, by construction, $P_\text{ram}$ and $\mathcal{M}_\text{3D}$ can be inferred with a markedly different precision. Within the inner regions of galaxy clusters,  the analytical ICM density profile is significantly steeper than the temperature profile. Consequently, the uncertainty in the determination of the 3D distance produces a larger scatter in $P_\text{ram}$ than in $\mathcal{M}_\text{3D}$. Additionally, uncertainties on $P_\text{ram}$ depends on the square of the uncertaintiy on the velocity, while it is linear for the Mach number. These effects lead to having larger uncertainties for $P_\text{ram}$ than for $\mathcal{M}_\text{3D}$, which is expected to have an impact on simple correlation tests. However, we stress that the analysis shown in Figure \ref{spear}, based on the full exploration of the Spearman rank distribution, is meant to bypass these problematics.
    \item At the current state, there is a series of parameters which are not taken into account into this method, such as disk-wind angle, ISM and ICM substructure, stripping evolutionary stage, ICM and ISM magnetic field orientation, and orbital interaction history. Exploring their impact requires tailored MHD simulations.
\end{itemize}

\section{Conclusions}
We presented new MeerKAT data to investigate the extraplanar magnetic field configuration in the galaxy JO147, finding it to differ substantially from the only other known case, JO206. We argue that this is because the two galaxies experenced different supersonic compressions during their orbit which result in different magnetic drape intensities where, in the case of JO206, the stronger compression enabled the extraplanar star formation. We then tested this scenario by comparing the Mach number and the extraplanar star formation of 17 GASP galaxies, finding evidence of a moderate positive correlation between these quantites. 

Therefore, we propose that the critical factor in forming a jellyfish galaxy could be the compression resulting from its supersonic orbits, which sets the conditions for both the stripped ISM and the magnetic field to harbor star formation outside the stellar disk. The Monte Carlo analysis suggests that the supersonic motion could be more impactful than the ram pressure, but we also note that disentangling the two effects is not trivial because they are both intrinsically dependent on the galaxy velocity. Furthermore, strong ram pressure events could have facilitated the removal of large gas clouds from the galaxy in the early stages of the stripping \citep[][]{Vollmer_2012}, which, due to the larger volume-to-surface ratio, would be more resilient to thermal evaporation than smaller ones \citep[][]{Armillotta_2017, Gronke2018}. Following our proposed framework, JO147 would exemplify a low-Mach-number stripping outcome, where the magnetic field configuration suggests a limited drape influence, and therefore the extraplanar star formation has been limited.

The connection between magnetic field, orbit velocity, and star formation that we have presented in this work can be further explored via numerical simulations, as already pioneered by \cite[][]{Sparre2020,Sparre2024}. Specifically, it is paramount to address the role of the inclination angle between the stellar disk and the wind, and the interplay between magnetic drape, star formation feedback, and ICM turbulent mixing \citep[][]{Franchetto_2021,Sun_2022,Ignesti_2024}. Furthermore, future wide-field, high-angular resolution X-ray observations with the AXIS telescope \citep[][]{AXIS_2025} may help confirming this scenario by providing direct cluster galaxies Mach number measurements by detecting the bow-shock induced by their supersonic motions. To further explore this scenario, it is now crucial to map the magnetic field of large cluster galaxy samples to estimate the incidence of ordered, large-scale field configurations via radio continuum observations. A complementary prediction of the proposed framework is that super-sonic galaxies, due to the higher magnetic field intensity in the magnetic drape and subsequent higher synchrotron emissivity, should show a larger fraction of polarized extraplanar emission associated with the ordered magnetic field than the other cluster galaxies. However, the extraplanar polarized emission is especially elusive in observations because of the combination of two competing processes, the spectral contraction of the radio continuum tails, whose extension typically drops above 1 GHz \citep[][]{Ignesti2022,Roberts_2024b}, and the ICM Faraday rotation, which grows with the ICM column density, hence damping the low-frequency polarized signal of those galaxies at the cluster center, hence more likely to develop stripped tails. These combined physical effects, on top of the typical low surface brightness of these sources, makes it difficult to find suitable candidates for these studies. Therefore, we expect that  only high-resolution polarimetry surveys between 1 and 2 GHz, such as the MeerKAT Fornax survey \citep[][]{Loi_2025} and, in the future, with the Square Kilometer Array \citep[][]{Loi_2019} in SKA-MID Band 2 can provide us with magnetic field topology for a large sample of cluster galaxies to further explore the physics regulating the evolution of the ISM under extreme conditions set by the ICM ram pressure. 
\begin{acknowledgements}
We acknowledge the constructive contribute of the Reviewers that improved the presentation of our work. Based on observations collected at the European Organization for Astronomical Research in the Southern Hemisphere under ESO programme 196.B-0578. This project has received funding from the European Research Council (ERC) under the European Union's Horizon 2020 research and innovation programme (grant agreement No. 833824). CP acknowledges support by the European Research Council under ERC-AdG grant PICOGAL-101019746. YLJ acknowledges support from the Agencia Nacional de Investigaci\'on y Desarrollo (ANID) through Basal project FB210003, FONDECYT Regular projects 1241426 and 123044, and  Millennium  Science Initiative Program NCN2024\_112. GP acknowledges funding from the European Union – NextGenerationEU, RRF M4C2 1.1, Project
2022JZJBHM: “AGN-sCAN: zooming-in on the AGN-galaxy connection since
the cosmic noon” – CUP C53D23001120006. PK is partially supported by the BMBF project 05A23PC1 for D-MeerKAT III. 
(Part of) the data published here have been reduced using the CARACal pipeline, partially supported by ERC Starting grant number 679627 “FORNAX”, MAECI Grant Number ZA18GR02, DST-NRF Grant Number 113121 as part of the ISARP Joint Research Scheme, and BMBF project 05A17PC2 for D-MeerKAT. Information about CARACal can be obtained online under the URL: https://caracal.readthedocs.io . A.E.L. and B.V. acknowledge support from the INAF GO grant 2023 ``Identifying ram pressure induced unwinding arms in cluster spirals'' (P.I. Vulcani). AI thanks the music of Foxy Shazam for inspiring the preparation of the draft.
\end{acknowledgements}

%

\bibliographystyle{aa}
\bibliography{sn-bibliography}%

@ARTICLE{Sparre2019,
       author = {{Sparre}, Martin and {Pfrommer}, Christoph and {Vogelsberger}, Mark},
        title = "{The physics of multiphase gas flows: fragmentation of a radiatively cooling gas cloud in a hot wind}",
      journal = {\mnras},
         year = 2019,
        month = feb,
       volume = {482},
       number = {4},
        pages = {5401-5421},
          doi = {10.1093/mnras/sty3063},
archivePrefix = {arXiv},
       eprint = {1807.07971},
 primaryClass = {astro-ph.GA},
       adsurl = {https://ui.adsabs.harvard.edu/abs/2019MNRAS.482.5401S}
}

@ARTICLE{Lee2022,
       author = {{Lee}, Jaehyun and {Kimm}, Taysun and {Blaizot}, J{\'e}r{\'e}my and {Katz}, Harley and {Lee}, Wonki and {Sheen}, Yun-Kyeong and {Devriendt}, Julien and {Slyz}, Adrianne},
        title = "{Simulating Jellyfish Galaxies: A Case Study for a Gas-rich Dwarf Galaxy}",
      journal = {\apj},
         year = 2022,
        month = apr,
       volume = {928},
       number = {2},
          eid = {144},
        pages = {144},
          doi = {10.3847/1538-4357/ac5595},
archivePrefix = {arXiv},
       eprint = {2201.01316},
 primaryClass = {astro-ph.GA},
       adsurl = {https://ui.adsabs.harvard.edu/abs/2022ApJ...928..144L}
}

@ARTICLE{Boselli2022,
       author = {{Boselli}, Alessandro and {Fossati}, Matteo and {Sun}, Ming},
        title = "{Ram pressure stripping in high-density environments}",
      journal = {\aapr},
         year = 2022,
        month = dec,
       volume = {30},
       number = {1},
          eid = {3},
        pages = {3},
          doi = {10.1007/s00159-022-00140-3},
archivePrefix = {arXiv},
       eprint = {2109.13614},
 primaryClass = {astro-ph.GA},
       adsurl = {https://ui.adsabs.harvard.edu/abs/2022A&ARv..30....3B}
}

@ARTICLE{Shapley_1930,
       author = {{Shapley}, Harlow},
        title = "{Note on a Remote Cloud of Galaxies in Centaurus}",
      journal = {Harvard College Observatory Bulletin},
         year = 1930,
        month = mar,
       volume = {874},
        pages = {9-12},
       adsurl = {https://ui.adsabs.harvard.edu/abs/1930BHarO.874....9S}
}

@ARTICLE{GdG_2025,
       author = {{Di Gennaro}, G. and {Venturi}, T. and {Giacintucci}, S. and {Br{\"u}ggen}, M. and {Bulbul}, E. and {Sanders}, J. and {Liu}, A. and {Zhang}, X. and {Trehaeven}, K. and {Dallacasa}, D. and {Merluzzi}, P. and {Pasini}, T. and {Bardelli}, S. and {Bernardi}, G. and {Smirnov}, O.},
        title = "{Cosmic dance in the Shapley Concentration Core: II. The uGMRT-MeerKAT view of filaments in the brightest cluster galaxies and tailed radio galaxies in the A3528 cluster complex}",
      journal = {\aap},
         year = 2025,
        month = feb,
       volume = {694},
          eid = {A28},
        pages = {A28},
          doi = {10.1051/0004-6361/202451066},
archivePrefix = {arXiv},
       eprint = {2408.14142},
 primaryClass = {astro-ph.CO},
       adsurl = {https://ui.adsabs.harvard.edu/abs/2025A&A...694A..28D}
}

@mastersthesis{Dusch2025,
  author      = {Niklas Dusch},
  title       = {The Impact of realistic galaxy orbits on jellyfish galaxy properties},
  type        = {masterthesis},
  pages       = {vi, 147},
  school      = {Universit{\"a}t Potsdam},
  year        = {2025},
  doi={https://doi.org/10.25932/publishup-69560}
}

@ARTICLE{Venturi_2000,
       author = {{Venturi}, T. and {Bardelli}, S. and {Morganti}, R. and {Hunstead}, R.~W.},
        title = "{Radio properties of the Shapley Concentration - III. Merging clusters in the A3558 complex}",
      journal = {\mnras},
         year = 2000,
        month = may,
       volume = {314},
       number = {3},
        pages = {594-610},
          doi = {10.1046/j.1365-8711.2000.03403.x},
archivePrefix = {arXiv},
       eprint = {astro-ph/0001256},
 primaryClass = {astro-ph},
       adsurl = {https://ui.adsabs.harvard.edu/abs/2000MNRAS.314..594V}
}

@ARTICLE{Merluzzi_2015,
       author = {{Merluzzi}, P. and {Busarello}, G. and {Haines}, C.~P. and {Mercurio}, A. and {Okabe}, N. and {Pimbblet}, K.~J. and {Dopita}, M.~A. and {Grado}, A. and {Limatola}, L. and {Bourdin}, H. and {Mazzotta}, P. and {Capaccioli}, M. and {Napolitano}, N.~R. and {Schipani}, P.},
        title = "{Shapley Supercluster Survey: Galaxy evolution from filaments to cluster cores}",
      journal = {\mnras},
         year = 2015,
        month = jan,
       volume = {446},
       number = {1},
        pages = {803-822},
          doi = {10.1093/mnras/stu2085},
archivePrefix = {arXiv},
       eprint = {1407.4628},
 primaryClass = {astro-ph.GA},
       adsurl = {https://ui.adsabs.harvard.edu/abs/2015MNRAS.446..803M}
}

@ARTICLE{Rossetti_2007,
       author = {{Rossetti}, M. and {Ghizzardi}, S. and {Molendi}, S. and {Finoguenov}, A.},
        title = "{A cluster in a crowded environment: XMM-Newton and Chandra observations of A3558}",
      journal = {\aap},
         year = 2007,
        month = mar,
       volume = {463},
       number = {3},
        pages = {839-851},
          doi = {10.1051/0004-6361:20054621},
archivePrefix = {arXiv},
       eprint = {astro-ph/0611056},
 primaryClass = {astro-ph},
       adsurl = {https://ui.adsabs.harvard.edu/abs/2007A&A...463..839R}
}

@ARTICLE{Venturi_2022,
       author = {{Venturi}, T. and {Giacintucci}, S. and {Merluzzi}, P. and {Bardelli}, S. and {Busarello}, G. and {Dallacasa}, D. and {Sikhosana}, S.~P. and {Marvil}, J. and {Smirnov}, O. and {Bourdin}, H. and {Mazzotta}, P. and {Rossetti}, M. and {Rudnick}, L. and {Bernardi}, G. and {Br{\"u}ggen}, M. and {Carretti}, E. and {Cassano}, R. and {Di Gennaro}, G. and {Gastaldello}, F. and {Kale}, R. and {Knowles}, K. and {Koribalski}, B.~S. and {Heywood}, I. and {Hopkins}, A.~M. and {Norris}, R.~P. and {Reiprich}, T.~H. and {Tasse}, C. and {Vernstrom}, T. and {Zucca}, E. and {Bester}, L.~H. and {Diego}, J.~M. and {Kanapathippillai}, J.},
        title = "{Radio footprints of a minor merger in the Shapley Supercluster: From supercluster down to galactic scales}",
      journal = {\aap},
         year = 2022,
        month = apr,
       volume = {660},
          eid = {A81},
        pages = {A81},
          doi = {10.1051/0004-6361/202142048},
archivePrefix = {arXiv},
       eprint = {2201.04887},
 primaryClass = {astro-ph.CO},
       adsurl = {https://ui.adsabs.harvard.edu/abs/2022A&A...660A..81V}
}

@ARTICLE{Vollmer_2004,
       author = {{Vollmer}, B. and {Thierbach}, M. and {Wielebinski}, R.},
        title = "{Radio continuum spectra of galaxies in the Virgo cluster region}",
      journal = {\aap},
         year = 2004,
        month = apr,
       volume = {418},
        pages = {1-6},
          doi = {10.1051/0004-6361:20035759},
archivePrefix = {arXiv},
       eprint = {astro-ph/0401104},
 primaryClass = {astro-ph},
       adsurl = {https://ui.adsabs.harvard.edu/abs/2004A&A...418....1V}
}

@ARTICLE{Vulcani_2015,
       author = {{Vulcani}, Benedetta and {Poggianti}, Bianca M. and {Fritz}, Jacopo and {Fasano}, Giovanni and {Moretti}, Alessia and {Calvi}, Rosa and {Paccagnella}, Angela},
        title = "{From Blue Star-forming to Red Passive: Galaxies in Transition in Different Environments}",
      journal = {\apj},
         year = 2015,
        month = jan,
       volume = {798},
       number = {1},
          eid = {52},
        pages = {52},
          doi = {10.1088/0004-637X/798/1/52},
archivePrefix = {arXiv},
       eprint = {1410.6481},
 primaryClass = {astro-ph.GA},
       adsurl = {https://ui.adsabs.harvard.edu/abs/2015ApJ...798...52V}
}

@ARTICLE{Li_2023,
       author = {{Li}, Yuan and {Luo}, Rongxin and {Fossati}, Matteo and {Sun}, Ming and {J{\'a}chym}, Pavel},
        title = "{Turbulence in the tail of a jellyfish galaxy}",
      journal = {\mnras},
         year = 2023,
        month = may,
       volume = {521},
       number = {3},
        pages = {4785-4791},
          doi = {10.1093/mnras/stad874},
archivePrefix = {arXiv},
       eprint = {2303.15500},
 primaryClass = {astro-ph.GA},
       adsurl = {https://ui.adsabs.harvard.edu/abs/2023MNRAS.521.4785L}
}

@ARTICLE{Roberts_2024b,
       author = {{Roberts}, I.~D. and {van Weeren}, R.~J. and {Lal}, D.~V. and {Sun}, M. and {Chen}, H. and {Ignesti}, A. and {Br{\"u}ggen}, M. and {Lyskova}, N. and {Venturi}, T. and {Yagi}, M.},
        title = "{Radio-continuum spectra of ram-pressure-stripped galaxies in the Coma Cluster}",
      journal = {\aap},
         year = 2024,
        month = mar,
       volume = {683},
          eid = {A11},
        pages = {A11},
          doi = {10.1051/0004-6361/202347977},
       adsurl = {https://ui.adsabs.harvard.edu/abs/2024A&A...683A..11R}
}

@ARTICLE{Vollmer_2012,
       author = {{Vollmer}, B. and {Soida}, M. and {Braine}, J. and {Abramson}, A. and {Beck}, R. and {Chung}, A. and {Crowl}, H.~H. and {Kenney}, J.~D.~P. and {van Gorkom}, J.~H.},
        title = "{Ram pressure stripping of the multiphase ISM and star formation in the Virgo spiral galaxy NGC 4330}",
      journal = {\aap},
         year = 2012,
        month = jan,
       volume = {537},
          eid = {A143},
        pages = {A143},
          doi = {10.1051/0004-6361/201117680},
archivePrefix = {arXiv},
       eprint = {1111.5236},
 primaryClass = {astro-ph.CO},
       adsurl = {https://ui.adsabs.harvard.edu/abs/2012A&A...537A.143V}
}

@ARTICLE{Serra_2024,
       author = {{Serra}, P. and {Oosterloo}, T.~A. and {Kamphuis}, P. and {J{\'o}zsa}, G.~I.~G. and {de Blok}, W.~J.~G. and {Bryan}, G.~L. and {van Gorkom}, J.~H. and {Iodice}, E. and {Kleiner}, D. and {Loni}, A. and {Loubser}, S.~I. and {Maccagni}, F.~M. and {Moln{\'a}r}, D. and {Peletier}, R. and {Pisano}, D.~J. and {Ramatsoku}, M. and {Smith}, M.~W.~L. and {Verheijen}, M.~A.~W. and {Zabel}, N.},
        title = "{The MeerKAT Fornax Survey: III. Ram-pressure stripping of the tidally interacting galaxy NGC 1427A in the Fornax cluster}",
      journal = {\aap},
         year = 2024,
        month = oct,
       volume = {690},
          eid = {A4},
        pages = {A4},
          doi = {10.1051/0004-6361/202450114},
archivePrefix = {arXiv},
       eprint = {2407.09082},
 primaryClass = {astro-ph.GA},
       adsurl = {https://ui.adsabs.harvard.edu/abs/2024A&A...690A...4S}
}

@ARTICLE{Sun_2025,
       author = {{Sun}, M. and {Le}, H. and {Epinat}, B. and {Boselli}, A. and {Luo}, R. and {Hosogi}, K. and {Pichette}, N. and {Forman}, W. and {Sarazin}, C. and {Fossati}, M. and {Chen}, H. and {Hensler}, G. and {Sarpa}, E. and {Amram}, P. and {Braine}, J. and {Cuillandre}, J.~C. and {Gwyn}, S. and {Martocchia}, S. and {Vollmer}, B.},
        title = "{A Virgo Environmental Survey Tracing Ionised Gas Emission (VESTIGE): XIX. The discovery of a spectacular 230 kpc H{\ensuremath{\alpha}} tail following NGC 4569 in the Virgo cluster}",
      journal = {\aap},
         year = 2026,
        month = jan,
       volume = {705},
          eid = {A139},
        pages = {A139},
          doi = {10.1051/0004-6361/202556225},
archivePrefix = {arXiv},
       eprint = {2507.02527},
 primaryClass = {astro-ph.GA},
       adsurl = {https://ui.adsabs.harvard.edu/abs/2026A&A...705A.139S}
}

@article{Gunn1972,
   author = {J.~E. Gunn and J.~R. Gott},
   doi = {10.1086/151605},
   month = {8},
   pages = {1-+},
   title = {On the Infall of Matter Into Clusters of Galaxies and Some Effects on Their Evolution},
   volume = {176},
    journal = {\apj},
   year = {1972}
}

@BOOK{Sarazin1988,
       author = {{Sarazin}, Craig L.},
        title = "{X-ray emission from clusters of galaxies}",
         year = 1988,
       adsurl = {https://ui.adsabs.harvard.edu/abs/1988xrec.book.....S}
}

@article{Poggianti2016,
   author = {B.~M. Poggianti and G Fasano and A Omizzolo and M Gullieuszik and D Bettoni and A Moretti and A Paccagnella and Y.~L. Jaffé and B Vulcani and J Fritz and W Couch and M D'Onofrio},
   doi = {10.3847/0004-6256/151/3/78},
   month = {3},
   pages = {78},
   journal = {\aj},
   title = {Jellyfish Galaxy Candidates at Low Redshift},
   volume = {151},
   year = {2016}
}

@article{Poggianti2017,
   author = {B.~M. Poggianti and A Moretti and M Gullieuszik and J Fritz and Y Jaffé and D Bettoni and G Fasano and C Bellhouse and G Hau and B Vulcani and A Biviano and A Omizzolo and A Paccagnella and M D'Onofrio and A Cava and Y.-K. Sheen and W Couch and M Owers},
   doi = {10.3847/1538-4357/aa78ed},
   month = {7},
   pages = {48},
   title = {GASP. I. Gas Stripping Phenomena in Galaxies with MUSE},
journal={\apj},
   volume = {844},
  journal = {\apj},
   year = {2017}
}

@article{Poggianti2019,
   author = {Bianca M Poggianti and Alessandro Ignesti and Myriam Gitti and Anna Wolter and Fabrizio Brighenti and Andrea Biviano and Koshy George and Benedetta Vulcani and Marco Gullieuszik and Alessia Moretti and Rosita Paladino and Daniela Bettoni and Andrea Franchetto and Yara L Jaffé and Mario Radovich and Elke Roediger and Neven Tomičić and Stephanie Tonnesen and Callum Bellhouse and Jacopo Fritz and Alessandro Omizzolo},
   doi = {10.3847/1538-4357/ab5224},
  journal = {\apj},
   issue = {2},
   month = {12},
   pages = {155},
   title = {GASP XXIII: A Jellyfish Galaxy as an Astrophysical Laboratory of the Baryonic Cycle},
   volume = {887},
   year = {2019}
}

@article{Sparre2024,
   author = {Martin Sparre and Christoph Pfrommer and Ewald Puchwein},
   doi = {10.1093/mnras/stad3607},
   issue = {3},
   month = {1},
   pages = {5829-5842},
  journal = {\mnras},
   title = {The magnetized and thermally unstable tails of jellyfish galaxies},
   volume = {527},
   year = {2024}
}

@ARTICLE{Sparre2024_CGM,
       author = {{Sparre}, Martin and {Pfrommer}, Christoph and {Puchwein}, Ewald},
        title = "{Comparing the interstellar and circumgalactic origin of gas in the tails of jellyfish galaxies}",
      journal = {\aap},
         year = 2024,
        month = nov,
       volume = {691},
          eid = {A259},
        pages = {A259},
          doi = {10.1051/0004-6361/202450544},
archivePrefix = {arXiv},
       eprint = {2405.00768},
 primaryClass = {astro-ph.GA},
       adsurl = {https://ui.adsabs.harvard.edu/abs/2024A&A...691A.259S}
}

@ARTICLE{Li2020,
       author = {{Li}, Zhihui and {Hopkins}, Philip F. and {Squire}, Jonathan and {Hummels}, Cameron},
        title = "{On the survival of cool clouds in the circumgalactic medium}",
      journal = {\mnras},
         year = 2020,
        month = feb,
       volume = {492},
       number = {2},
        pages = {1841-1854},
          doi = {10.1093/mnras/stz3567},
archivePrefix = {arXiv},
       eprint = {1909.02632},
 primaryClass = {astro-ph.GA},
       adsurl = {https://ui.adsabs.harvard.edu/abs/2020MNRAS.492.1841L}
}

@article{Dursi2008,
   author = {L.~J. Dursi and C Pfrommer},
   doi = {10.1086/529371},
   issue = {2},
   month = {4},
   pages = {993-1018},
    journal = {\apj},
   title = {Draping of Cluster Magnetic Fields over Bullets and BubblesMorphology and Dynamic Effects},
   volume = {677},
   year = {2008}
}

@article{Pfrommer2010,
   author = {Christoph Pfrommer and L Jonathan Dursi},
   doi = {10.1038/nphys1657},
   issue = {7},
   journal = {Nature Physics},
   month = {7},
   pages = {520-526},
   title = {Detecting the orientation of magnetic fields in galaxy clusters},
   volume = {6},
   year = {2010}
}

@article{Govoni2004,
   author = {F Govoni and L Feretti},
   doi = {10.1142/S0218271804005080},
   journal = {International Journal of Modern Physics D},
   pages = {1549-1594},
   title = {Magnetic Fields in Clusters of Galaxies},
   volume = {13},
   year = {2004}
}

@article{Mller2021,
   author = {Ancla Müller and Bianca Maria Poggianti and Christoph Pfrommer and Björn Adebahr and Paolo Serra and Alessandro Ignesti and Martin Sparre and Myriam Gitti and Ralf-Jürgen Dettmar and Benedetta Vulcani and Alessia Moretti},
   doi = {10.1038/s41550-020-01234-7},
   journal = {Nature Astronomy},
   month = {1},
   pages = {159-168},
   title = {Highly ordered magnetic fields in the tail of the jellyfish galaxy JO206},
   volume = {5},
   year = {2021}
}

@article{Sparre2020,
   author = {Martin Sparre and Christoph Pfrommer and Kristian Ehlert},
   doi = {10.1093/mnras/staa3177},
   issue = {3},
    journal = {\mnras},
   month = {10},
   pages = {4261-4281},
   title = {Interaction of a cold cloud with a hot wind: the regimes of cloud growth and destruction and the impact of magnetic fields},
   volume = {499},
   year = {2020}
}

@article{Poggianti_2019,
   author = {Bianca M Poggianti and Marco Gullieuszik and Stephanie Tonnesen and Alessia Moretti and Benedetta Vulcani and Mario Radovich and Yara Jaffé and Jacopo Fritz and Daniela Bettoni and Andrea Franchetto and Giovanni Fasano and Callum Bellhouse and Alessandro Omizzolo},
   doi = {10.1093/mnras/sty2999},
   journal = {\mnras},
   month = {2},
   pages = {4466-4502},
   title = {GASP XIII. Star formation in gas outside galaxies},
   volume = {482},
   year = {2019}
}

@article{Ignesti2023,
   author = {A Ignesti and B Vulcani and A Botteon and B Poggianti and E Giunchi and R Smith and G Brunetti and I.~D. Roberts and R.~J. van Weeren and K Rajpurohit},
   doi = {10.1051/0004-6361/202346517},
    journal = {\aap},
   month = {7},
   pages = {A118},
   title = {Radio continuum tails in ram pressure-stripped spiral galaxies: Experimenting with a semi-empirical model in Abell 2255},
   volume = {675},
   year = {2023}
}

@article{Gullieuszik2020,
   author = {Marco Gullieuszik and Bianca M Poggianti and Sean L McGee and Alessia Moretti and Benedetta Vulcani and Stephanie Tonnesen and Elke Roediger and Yara L Jaffé and Jacopo Fritz and Andrea Franchetto and Alessandro Omizzolo and Daniela Bettoni and Mario Radovich and Anna Wolter},
   doi = {10.3847/1538-4357/aba3cb},
  journal = {\apj},
   issue = {1},
   month = {8},
   pages = {13},
   title = {GASP. XXI. Star Formation Rates in the Tails of Galaxies Undergoing Ram Pressure Stripping},
   volume = {899},
   year = {2020}
}

@article{Bardelli1996,
   author = {S Bardelli and E Zucca and A Malizia and G Zamorani and R Scaramella and G Vettolani},
   doi = {10.48550/arXiv.astro-ph/9506019},
  journal = {\aap},
   month = {1},
   pages = {435},
   title = {A study of the core of the Shapley concentration. II. ROSAT observation of A 3558.},
   volume = {305},
   year = {1996}
}

@article{Smith2022b,
   author = {Rory Smith and Jong-Ho Shinn and Stephanie Tonnesen and Paula Calderón-Castillo and Jacob Crossett and Yara L Jaffe and Ian Roberts and Sean McGee and Koshy George and Benedetta Vulcani and Marco Gullieuszik and Alessia Moretti and Bianca M Poggianti and Jihye Shin},
   doi = {10.3847/1538-4357/ac7ab5},
     journal = {\apj},
   issue = {1},
   month = {7},
   pages = {86},
   title = {A New Method to Constrain the Appearance and Disappearance of Observed Jellyfish Galaxy Tails},
   volume = {934},
   year = {2022}
}

@article{Gronke2018,
   author = {Max Gronke and S Peng Oh},
   doi = {10.1093/mnrasl/sly131},
      journal = {\mnras},
   issue = {1},
   month = {10},
   pages = {L111-L115},
   title = {The growth and entrainment of cold gas in a hot wind},
   volume = {480},
   year = {2018}
}

@INPROCEEDINGS{Jonas_2016,
       author = {{Jonas}, J. and {MeerKAT Team}},
        title = "{The MeerKAT Radio Telescope}",
    booktitle = {MeerKAT Science: On the Pathway to the SKA},
         year = 2016,
        month = jan,
          eid = {1},
        pages = {1},
          doi = {10.22323/1.277.0001},
       adsurl = {https://ui.adsabs.harvard.edu/abs/2016mks..confE...1J}
}

@article{Biviano2017,
   author = {A Biviano and A Moretti and A Paccagnella and B.~M. Poggianti and D Bettoni and M Gullieuszik and B Vulcani and G Fasano and M D'Onofrio and J Fritz and A Cava},
   doi = {10.1051/0004-6361/201731289},
   journal = {\aap},
   month = {11},
   pages = {A81},
   title = {The concentration-mass relation of clusters of galaxies from the OmegaWINGS survey},
   volume = {607},
   year = {2017}
}

@article{Smith2022,
   author = {Rory Smith and Jong-Ho Shinn and Stephanie Tonnesen and Paula Calderón-Castillo and Jacob Crossett and Yara L Jaffe and Ian Roberts and Sean McGee and Koshy George and Benedetta Vulcani and Marco Gullieuszik and Alessia Moretti and Bianca M Poggianti and Jihye Shin},
   doi = {10.3847/1538-4357/ac7ab5},
      journal = {\apj},
   issue = {1},
   month = {7},
   pages = {86},
   title = {A New Method to Constrain the Appearance and Disappearance of Observed Jellyfish Galaxy Tails},
   volume = {934},
   year = {2022}
}

@article{gala,
  doi = {10.21105/joss.00388},
  url = {https://doi.org/10.21105%2Fjoss.00388},
  year = 2017,
  month = {oct},
  publisher = {The Open Journal},
  volume = {2},
  number = {18},
  author = {Adrian M. Price-Whelan},
  title = {Gala: A Python package for galactic dynamics},
  journal = {The Journal of Open Source Software}}

@ARTICLE{Vulcani_2018,
       author = {{Vulcani}, Benedetta and {Poggianti}, Bianca M. and {Gullieuszik}, Marco and {Moretti}, Alessia and {Tonnesen}, Stephanie and {Jaff{\'e}}, Yara L. and {Fritz}, Jacopo and {Fasano}, Giovanni and {Bettoni}, Daniela},
        title = "{Enhanced Star Formation in Both Disks and Ram-pressure-stripped Tails of GASP Jellyfish Galaxies}",
      journal = {\apjl},
         year = 2018,
        month = oct,
       volume = {866},
       number = {2},
          eid = {L25},
        pages = {L25},
          doi = {10.3847/2041-8213/aae68b},
archivePrefix = {arXiv},
       eprint = {1810.05164},
 primaryClass = {astro-ph.GA},
       adsurl = {https://ui.adsabs.harvard.edu/abs/2018ApJ...866L..25V}
}

@ARTICLE{AXIS_2025,
       author = {{Koss}, Michael and {Aftab}, Nafisa and {Allen}, Steven W. and {Amato}, Roberta and {An}, Hongjun and {Andreoni}, Igor and {Anguita}, Timo and {Arcodia}, Riccardo and {Ayres}, Thomas and {Bachetti}, Matteo and {Baglio}, Maria Cristina and {Bahramian}, Arash and {Balboni}, Marco and {Baldi}, Ranieri D. and {Balman}, Solen and {Bamba}, Aya and {Banados}, Eduardo and {Bao}, Tong and {Bartalucci}, Iacopo and {Basu-Zych}, Antara and {Batalha}, Rebeca and {Battistini}, Lorenzo and {Bauer}, Franz Erik and {Beardmore}, Andy and {Becker}, Werner and {Behar}, Ehud and {Belfiore}, Andrea and {Beniamini}, Paz and {Bertola}, Elena and {Bessa}, Vinicius and {Best}, Henry and {Bianchi}, Stefano and {Biava}, N. and {Binder}, Breanna A. and {Blanton}, Elizabeth L. and {Bodaghee}, Arash and {Bogdanovic}, Tamara and {Bogensberger}, David and {Bonafede}, A. and {Bonetti}, Matteo and {Bordas}, Pol and {Borghese}, Alice and {Botteon}, Andrea and {Boula}, Stella and {Bozzo}, Enrico and {Branchesi}, Marica and {Brandt}, William Nielsen and {Bregman}, Joel and {Brighenti}, Fabrizio and {Bronzini}, Ettore and {Brunelli}, Giulia and {Brusa}, Marcella and {Bulbul}, Esra and {Burdge}, Kevin and {Caccianiga}, Alessandro and {Calzadilla}, Michael and {Campana}, Sergio and {Capalbi}, Milvia and {Capitanio}, Fiamma and {Cappelluti}, Nico and {Carney}, Jonathan and {Casanova}, Sabrina and {Castro}, Daniel and {Cenko}, S. Bradley and {Chakraborty}, Joheen and {Chakraborty}, Priyanka and {Chartas}, George and {Chatterjee}, Arka and {Choudhury}, Prakriti Pal and {Cilley}, Raven and {Civano}, Francesca and {Comastri}, Andrea and {Connor}, Thomas and {Corcoran}, Michael F. and {Corrales}, Lia and {Coti Zelati}, Francesco and {Cui}, Weiguang and {D'Ammando}, Filippo and {Dage}, Kristen and {Daylan}, Tansu and {De Grandi}, Sabrina and {De Rosa}, Alessandra and {Decarli}, Roberto and {Decourchelle}, Anne and {Degenaar}, Nathalie and {Del Popolo}, Antonino and {Di Marco}, Alessandro and {Di Salvo}, Tiziana and {Dichiara}, Simone and {DiKerby}, Stephen and {Dillmann}, Steven and {Doerksen}, Neil and {Draghis}, Paul and {Drake}, Jeremy J. and {Ducci}, Lorenzo and {Dupke}, Renato and {Durbak}, Joseph and {Duvvuri}, Girish M. and {Dykaar}, Hannah and {Eckert}, Dominique and {Elvis}, Martin and {Espaillat}, Catherine and {Esposito}, Paolo and {Furst}, Felix and {Fabbiano}, Giuseppina and {Fagin}, Joshua and {Falcone}, Abraham and {Fedorova}, Elena and {Feinstein}, Adina and {Fernandez Fernandez}, Jorge and {Ferrand}, Gilles and {Flores}, Anthony M. and {Foo}, N. and {Foo}, Nicholas and {Foord}, Adi and {Franchini}, Alessia and {Fraschetti}, Federico and {Frye}, Brenda L. and {Lowenthal}, James D. and {Fryer}, Chris and {Fujimoto}, Shin-ichiro and {Gagnon}, Seth and {Gallo}, Luigi and {Garcia Diaz}, Carlos and {Gaspari}, Massimo and {Gastaldello}, Fabio and {Gelfand}, Joseph D. and {Gezari}, Suvi and {Ghizzardi}, Simona and {Giacintucci}, Simona and {Gill}, A. and {Gilli}, Roberto and {Gitti}, Myriam and {Giustini}, Margherita and {Gnarini}, Andrea and {Grandi}, Paola and {Gross}, Arran and {Gu}, Liyi and {Gunderson}, Sean and {Gunther}, Hans Moritz and {Haggard}, Daryl and {Hamaguchi}, Kenji and {Hare}, Jeremy and {Harrington}, Kevin C. and {Heinke}, Craig and {Heinz}, Sebastian and {Hlavacek-Larrondo}, Julie and {Ho}, Wynn C.~G. and {Hodges-Kluck}, Edmund and {Homan}, Jeroen and {Huang}, R. and {Ighina}, Luca and {Ignesti}, Alessandro and {Imbrogno}, Matteo and {Irwin}, Christopher and {Irwin}, Jimmy and {Islam}, Nazma and {Israel}, Gian Luca and {Jacobson-Galan}, Wynn and {Jain}, Chetana and {Jana}, Arghajit and {Jaodand}, Amruta and {Jennings}, Fred and {Jiang}, Jiachen and {Jimenez-Andrade}, Eric F. and {Jimenez-Teja}, Y. and {Johnson}, S. and {Jonker}, Peter and {Kamieneski}, Patrick S. and {Kammoun}, Elias and {Kara}, Erin and {Kargaltsev}, Oleg and {King}, George W. and {Kirmizibayrak}, Demet and {Klingler}, Noel and {Kong}, Albert K.~H. and {Kounkel}, Marina and {Kumar}, Manish and {Kutyrev}, Alexander and {Kyer}, Rebecca and {La Monaca}, Fabio and {Lambrides}, Erini and {Lanzuisi}, Giorgio and {Lee}, Wonki and {Lehmer}, Bret and {Lentini}, Elisa and {Lepore}, Marika and {Li}, Jiangtao and {Lisse}, Carey M. and {Liu}, Daizhong and {Liu}, Tingting and {Isla Llave}, Monica and {Locatelli}, Nicola and {Lopez}, Laura A. and {Lopez}, Sebastian and {Lovisari}, Lorenzo and {Lusso}, Elisabeta and {Mac Intyre}, Brydyn and {MacMaster}, Austin and {Maiolino}, Roberto},
        title = "{The Advanced X-ray Imaging Satellite Community Science Book}",
      journal = {arXiv e-prints},
         year = 2025,
        month = oct,
          eid = {arXiv:2511.00253},
        pages = {arXiv:2511.00253},
          doi = {10.48550/arXiv.2511.00253},
archivePrefix = {arXiv},
       eprint = {2511.00253},
 primaryClass = {astro-ph.HE},
       adsurl = {https://ui.adsabs.harvard.edu/abs/2025arXiv251100253K}
}

@ARTICLE{Loi_2019,
       author = {{Loi}, F. and {Murgia}, M. and {Govoni}, F. and {Vacca}, V. and {Prandoni}, I. and {Bonafede}, A. and {Feretti}, L.},
        title = "{Simulations of the polarized radio sky and predictions on the confusion limit in polarization for future radio surveys}",
      journal = {\mnras},
         year = 2019,
        month = jun,
       volume = {485},
       number = {4},
        pages = {5285-5293},
          doi = {10.1093/mnras/stz350},
archivePrefix = {arXiv},
       eprint = {1902.05953},
 primaryClass = {astro-ph.GA},
       adsurl = {https://ui.adsabs.harvard.edu/abs/2019MNRAS.485.5285L}
}

@ARTICLE{Loi_2025,
       author = {{Loi}, F. and {Serra}, P. and {Murgia}, M. and {Govoni}, F. and {Vacca}, V. and {Maccagni}, F. and {Kleiner}, D. and {Kamphuis}, P.},
        title = "{The MeerKAT Fornax Survey: IV. A close look at the cluster physics through the densest rotation measure grid}",
      journal = {\aap},
         year = 2025,
        month = feb,
       volume = {694},
          eid = {A125},
        pages = {A125},
          doi = {10.1051/0004-6361/202451711},
archivePrefix = {arXiv},
       eprint = {2501.05519},
 primaryClass = {astro-ph.CO},
       adsurl = {https://ui.adsabs.harvard.edu/abs/2025A&A...694A.125L}
}

@ARTICLE{Vikhlinin2006,
       author = {{Vikhlinin}, A. and {Kravtsov}, A. and {Forman}, W. and {Jones}, C. and {Markevitch}, M. and {Murray}, S.~S. and {Van Speybroeck}, L.},
        title = "{Chandra Sample of Nearby Relaxed Galaxy Clusters: Mass, Gas Fraction, and Mass-Temperature Relation}",
      journal = {\apj},
         year = 2006,
        month = apr,
       volume = {640},
       number = {2},
        pages = {691-709},
          doi = {10.1086/500288},
archivePrefix = {arXiv},
       eprint = {astro-ph/0507092},
 primaryClass = {astro-ph},
       adsurl = {https://ui.adsabs.harvard.edu/abs/2006ApJ...640..691V}
}

@ARTICLE{Biviano_2024,
       author = {{Biviano}, Andrea and {Poggianti}, Bianca M. and {Jaff{\'e}}, Yara and {Louren{\c{c}}o}, Ana C. and {Pizzuti}, Lorenzo and {Moretti}, Alessia and {Vulcani}, Benedetta},
        title = "{The Radial Orbits of Ram-pressure-stripped Galaxies in Clusters from the GASP Survey}",
      journal = {\apj},
         year = 2024,
        month = apr,
       volume = {965},
       number = {2},
          eid = {117},
        pages = {117},
          doi = {10.3847/1538-4357/ad2c09},
archivePrefix = {arXiv},
       eprint = {2403.02111},
 primaryClass = {astro-ph.CO},
       adsurl = {https://ui.adsabs.harvard.edu/abs/2024ApJ...965..117B}
}

@article{Ignesti2022,
   author = {Alessandro Ignesti and Benedetta Vulcani and Bianca M Poggianti and Rosita Paladino and Timothy Shimwell and Julia Healy and Myriam Gitti and Cecilia Bacchini and Alessia Moretti and Mario Radovich and Reinout J van Weeren and Ian D Roberts and Andrea Botteon and Ancla Müller and Sean McGee and Jacopo Fritz and Neven Tomičić and Ariel Werle and Matilde Mingozzi and Marco Gullieuszik and Marc Verheijen},
   doi = {10.3847/1538-4357/ac32ce},
     journal = {\apj},
   issue = {2},
   month = {1},
   pages = {64},
   title = {GASP XXXVIII: The LOFAR-MeerKAT-VLA View on the Nonthermal Side of a Jellyfish Galaxy},
   volume = {924},
   year = {2022}
}

@article{Navarro1997,
   author = {Julio F Navarro and Carlos S Frenk and Simon D.~M. White},
   doi = {10.1086/304888},
      journal = {\apj},
   issue = {2},
   month = {12},
   pages = {493-508},
   title = {A Universal Density Profile from Hierarchical Clustering},
   volume = {490},
   year = {1997}
}

@ARTICLE{Vollmer_2024,
       author = {{Vollmer}, B. and {Sun}, M. and {Jachym}, P. and {Fossati}, M. and {Boselli}, A.},
        title = "{ESO 137{\textendash}001: A jellyfish galaxy model}",
      journal = {\aap},
         year = 2024,
        month = dec,
       volume = {692},
          eid = {A4},
        pages = {A4},
          doi = {10.1051/0004-6361/202450435},
archivePrefix = {arXiv},
       eprint = {2409.16846},
 primaryClass = {astro-ph.GA},
       adsurl = {https://ui.adsabs.harvard.edu/abs/2024A&A...692A...4V}
}

@ARTICLE{Roberts_2024,
       author = {{Roberts}, I.~D. and {van Weeren}, R.~J. and {de Gasperin}, F. and {Botteon}, A. and {Edler}, H.~W. and {Ignesti}, A. and {Matijevi{\'c}}, L. and {Tomi{\v{c}}i{\'c}}, N.},
        title = "{A 100 kpc ram pressure tail trailing the group galaxy NGC 2276}",
      journal = {\aap},
         year = 2024,
        month = sep,
       volume = {689},
          eid = {A22},
        pages = {A22},
          doi = {10.1051/0004-6361/202450672},
archivePrefix = {arXiv},
       eprint = {2406.09221},
 primaryClass = {astro-ph.GA},
       adsurl = {https://ui.adsabs.harvard.edu/abs/2024A&A...689A..22R}
}

@INPROCEEDINGS{jozsa2020,
       author = {{J{\'o}zsa}, G.~I.~G. and {White}, S.~V. and {Thorat}, K. and {Smirnov}, O.~M. and {Serra}, P. and {Ramatsoku}, M. and {Ramaila}, A.~J.~T. and {Perkins}, S.~J. and {Maccagni}, F.~M. and {Makhathini}, S. and {Moln{\'a}r}, D.~C. and {Kamphuis}, P. and {Kleiner}, D. and {Hugo}, B.~V. and {de Blok}, W.~J.~G. and {Andati}, L.~A.~L.},
        title = "{MeerKATHI - an End-to-End Data Reduction Pipeline for MeerKAT and Other Radio Telescopes}",
    booktitle = {Astronomical Data Analysis Software and Systems XXIX},
         year = 2020,
       editor = {{Pizzo}, R. and {Deul}, E.~R. and {Mol}, J.~D. and {de Plaa}, J. and {Verkouter}, H.},
       series = {Astronomical Society of the Pacific Conference Series},
       volume = {527},
        month = jan,
        pages = {635},
          doi = {10.48550/arXiv.2006.02955},
archivePrefix = {arXiv},
       eprint = {2006.02955},
 primaryClass = {astro-ph.IM},
       adsurl = {https://ui.adsabs.harvard.edu/abs/2020ASPC..527..635J}
}

@ARTICLE{offringa2012,
       author = {{Offringa}, A.~R. and {van de Gronde}, J.~J. and {Roerdink}, J.~B.~T.~M.},
        title = "{A morphological algorithm for improving radio-frequency interference detection}",
      journal = {\aap},
         year = 2012,
        month = mar,
       volume = {539},
          eid = {A95},
        pages = {A95},
          doi = {10.1051/0004-6361/201118497},
archivePrefix = {arXiv},
       eprint = {1201.3364},
 primaryClass = {astro-ph.IM},
       adsurl = {https://ui.adsabs.harvard.edu/abs/2012A&A...539A..95O}
}

@ARTICLE{perley2017,
       author = {{Perley}, R.~A. and {Butler}, B.~J.},
        title = "{An Accurate Flux Density Scale from 50 MHz to 50 GHz}",
      journal = {\apjs},
         year = 2017,
        month = may,
       volume = {230},
       number = {1},
          eid = {7},
        pages = {7},
          doi = {10.3847/1538-4365/aa6df9},
archivePrefix = {arXiv},
       eprint = {1609.05940},
 primaryClass = {astro-ph.IM},
       adsurl = {https://ui.adsabs.harvard.edu/abs/2017ApJS..230....7P}
}

@ARTICLE{serra2015,
       author = {{Serra}, Paolo and {Westmeier}, Tobias and {Giese}, Nadine and {Jurek}, Russell and {Fl{\"o}er}, Lars and {Popping}, Attila and {Winkel}, Benjamin and {van der Hulst}, Thijs and {Meyer}, Martin and {Koribalski}, B{\"a}rbel S. and {Staveley-Smith}, Lister and {Courtois}, H{\'e}l{\`e}ne},
        title = "{SOFIA: a flexible source finder for 3D spectral line data}",
      journal = {\mnras},
         year = 2015,
        month = apr,
       volume = {448},
       number = {2},
        pages = {1922-1929},
          doi = {10.1093/mnras/stv079},
archivePrefix = {arXiv},
       eprint = {1501.03906},
 primaryClass = {astro-ph.IM},
       adsurl = {https://ui.adsabs.harvard.edu/abs/2015MNRAS.448.1922S}
}

@MISC{rmtools,
       author = {{Purcell}, C.~R. and {Van Eck}, C.~L. and {West}, J. and {Sun}, X.~H. and {Gaensler}, B.~M.},
        title = "{RM-Tools: Rotation measure (RM) synthesis and Stokes QU-fitting}",
 howpublished = {Astrophysics Source Code Library, record ascl:2005.003},
         year = 2020,
        month = may,
          eid = {ascl:2005.003},
        pages = {ascl:2005.003},
archivePrefix = {ascl},
       eprint = {2005.003},
       adsurl = {https://ui.adsabs.harvard.edu/abs/2020ascl.soft05003P}
}

@ARTICLE{george2012,
       author = {{George}, Samuel J. and {Stil}, Jeroen M. and {Keller}, Ben W.},
        title = "{Detection Thresholds and Bias Correction in Polarized Intensity}",
      journal = {\pasa},
         year = 2012,
        month = oct,
       volume = {29},
       number = {3},
        pages = {214-220},
          doi = {10.1071/AS11027},
archivePrefix = {arXiv},
       eprint = {1106.5362},
 primaryClass = {astro-ph.IM},
       adsurl = {https://ui.adsabs.harvard.edu/abs/2012PASA...29..214G}
}

@ARTICLE{2024MNRAS.533.1394M,
       author = {{Merluzzi}, P. and {Venturi}, T. and {Busarello}, G. and {Gennaro}, G. Di and {Giacintucci}, S. and {Casasola}, V. and {Krajnovi{\'c}}, D. and {Vernstrom}, T. and {Carretti}, E. and {Smirnov}, O. and {Trehaeven}, K. and {Anderson}, C.~S. and {Chesters}, J. and {Heald}, G. and {Hopkins}, A.~M. and {Koribalski}, B.},
        title = "{Ram-pressure stripped radio tails detected in the dynamically active environment of the Shapley Supercluster}",
      journal = {\mnras},
         year = 2024,
        month = sep,
       volume = {533},
       number = {2},
        pages = {1394-1411},
          doi = {10.1093/mnras/stae1867},
archivePrefix = {arXiv},
       eprint = {2407.21628},
 primaryClass = {astro-ph.GA},
       adsurl = {https://ui.adsabs.harvard.edu/abs/2024MNRAS.533.1394M}
}

@article{Cavagnolo2009,
   author = {K.~W. Cavagnolo and M Donahue and G.~M. Voit and M Sun},
   doi = {10.1088/0067-0049/182/1/12},
  journal = {\apjs},
   month = {5},
   pages = {12-32},
   title = {Intracluster Medium Entropy Profiles for a Chandra Archival Sample of Galaxy Clusters},
   volume = {182},
   year = {2009}
}

@ARTICLE{Sanderson2006,
       author = {{Sanderson}, Alastair J.~R. and {Ponman}, Trevor J. and {O'Sullivan}, Ewan},
        title = "{A statistically selected Chandra sample of 20 galaxy clusters - I. Temperature and cooling time profiles}",
      journal = {\mnras},
         year = 2006,
        month = nov,
       volume = {372},
       number = {4},
        pages = {1496-1508},
          doi = {10.1111/j.1365-2966.2006.10956.x},
archivePrefix = {arXiv},
       eprint = {astro-ph/0608423},
 primaryClass = {astro-ph},
       adsurl = {https://ui.adsabs.harvard.edu/abs/2006MNRAS.372.1496S}
}

@article{Bulbul2024,
   author = {E Bulbul and A Liu and M Kluge and X Zhang and J.~S. Sanders and Y.~E. Bahar and V Ghirardini and E Artis and R Seppi and C Garrel and M.~E. Ramos-Ceja and J Comparat and F Balzer and K Böckmann and M Brüggen and N Clerc and K Dennerl and K Dolag and M Freyberg and S Grandis and D Gruen and F Kleinebreil and S Krippendorf and G Lamer and A Merloni and K Migkas and K Nandra and F Pacaud and P Predehl and T.~H. Reiprich and T Schrabback and A Veronica and J Weller and S Zelmer},
   doi = {10.1051/0004-6361/202348264},
      journal = {\aap},
   month = {5},
   pages = {A106},
   title = {The SRG/eROSITA All-Sky Survey. The first catalog of galaxy clusters and groups in the Western Galactic Hemisphere},
   volume = {685},
   year = {2024}
}

@article{Bartolini2022,
   author = {Chiara Bartolini and Alessandro Ignesti and Myriam Gitti and Fabrizio Brighenti and Anna Wolter and Alessia Moretti and Benedetta Vulcani and Bianca M Poggianti and Marco Gullieuszik and Jacopo Fritz and Neven Tomičić},
   doi = {10.3847/1538-4357/ac866a},
    journal = {\apj},
   issue = {1},
   month = {9},
   pages = {74},
   title = {Unveiling the Interplay between the GASP Jellyfish Galaxy JO194 and Its Environment with Chandra},
   volume = {936},
   year = {2022}
}

@ARTICLE{David_1993,
       author = {{David}, L.~P. and {Slyz}, A. and {Jones}, C. and {Forman}, W. and {Vrtilek}, S.~D. and {Arnaud}, K.~A.},
        title = "{A Catalog of Intracluster Gas Temperatures}",
      journal = {\apj},
         year = 1993,
        month = aug,
       volume = {412},
        pages = {479},
          doi = {10.1086/172936},
       adsurl = {https://ui.adsabs.harvard.edu/abs/1993ApJ...412..479D}
}

@article{Hester2006,
   author = {J.~A. Hester},
   doi = {10.1086/505614},
    journal = {\apj},
   month = {8},
   pages = {910-921},
   title = {Ram Pressure Stripping in Clusters and Groups},
   volume = {647},
   year = {2006}
}

@BOOK{Spitzer_1978,
       author = {{Spitzer}, Lyman},
        title = "{Physical processes in the interstellar medium}",
         year = 1978,
          doi = {10.1002/9783527617722},
       adsurl = {https://ui.adsabs.harvard.edu/abs/1978ppim.book.....S}
}

@ARTICLE{Yun_2019,
       author = {{Yun}, Kiyun and {Pillepich}, Annalisa and {Zinger}, Elad and {Nelson}, Dylan and {Donnari}, Martina and {Joshi}, Gandhali and {Rodriguez-Gomez}, Vicente and {Genel}, Shy and {Weinberger}, Rainer and {Vogelsberger}, Mark and {Hernquist}, Lars},
        title = "{Jellyfish galaxies with the IllustrisTNG simulations - I. Gas-stripping phenomena in the full cosmological context}",
      journal = {\mnras},
         year = 2019,
        month = feb,
       volume = {483},
       number = {1},
        pages = {1042-1066},
          doi = {10.1093/mnras/sty3156},
archivePrefix = {arXiv},
       eprint = {1810.00005},
 primaryClass = {astro-ph.GA},
       adsurl = {https://ui.adsabs.harvard.edu/abs/2019MNRAS.483.1042Y}
}

@ARTICLE{Ebeling_2014,
       author = {{Ebeling}, H. and {Stephenson}, L.~N. and {Edge}, A.~C.},
        title = "{Jellyfish: Evidence of Extreme Ram-pressure Stripping in Massive Galaxy Clusters}",
      journal = {\apjl},
         year = 2014,
        month = feb,
       volume = {781},
       number = {2},
          eid = {L40},
        pages = {L40},
          doi = {10.1088/2041-8205/781/2/L40},
archivePrefix = {arXiv},
       eprint = {1312.6135},
 primaryClass = {astro-ph.GA},
       adsurl = {https://ui.adsabs.harvard.edu/abs/2014ApJ...781L..40E}
}

@article{Smith2010,
   author = {R.~J. Smith and J.~R. Lucey and D Hammer and A.~E. Hornschemeier and D Carter and M.~J. Hudson and R.~O. Marzke and M Mouhcine and S Eftekharzadeh and P James and H Khosroshahi and E Kourkchi and A Karick},
   doi = {10.1111/j.1365-2966.2010.17253.x},
      journal = {\mnras},
   month = {11},
   pages = {1417-1432},
   title = {Ultraviolet tails and trails in cluster galaxies: a sample of candidate gaseous stripping events in Coma},
   volume = {408},
   year = {2010}
}

@article{Ignesti2022b,
   author = {A Ignesti},
   doi = {10.1016/j.newast.2021.101732},
      journal = {\na},
   month = {4},
   pages = {101732},
   title = {Introducing PT-REX, the point-to-point TRend EXtractor},
   volume = {92},
   year = {2022}
}

@article{Klein1994,
   author = {Richard I Klein and Christopher F McKee and Philip Colella},
   doi = {10.1086/173554},
      journal = {\apj},
   month = {1},
   pages = {213},
   title = {On the Hydrodynamic Interaction of Shock Waves with Interstellar Clouds. I. Nonradiative Shocks in Small Clouds},
   volume = {420},
   year = {1994}
}

@ARTICLE{Babyk2023,
       author = {{Babyk}, Iurii V. and {McNamara}, Brian R.},
        title = "{The Halo Mass-Temperature Relation for Clusters, Groups, and Galaxies}",
      journal = {\apj},
         year = 2023,
        month = mar,
       volume = {946},
       number = {1},
          eid = {54},
        pages = {54},
          doi = {10.3847/1538-4357/acbf4b},
archivePrefix = {arXiv},
       eprint = {2302.11247},
 primaryClass = {astro-ph.GA},
       adsurl = {https://ui.adsabs.harvard.edu/abs/2023ApJ...946...54B}
}

@ARTICLE{Waldron2023,
       author = {{Waldron}, William and {Sun}, Ming and {Luo}, Rongxin and {Laudari}, Sunil and {Chatzikos}, Marios and {Sivanandam}, Suresh and {Kenney}, Jeffrey D.~P. and {J{\'a}chym}, Pavel and {Voit}, G. Mark and {Donahue}, Megan and {Fossati}, Matteo},
        title = "{HST viewing of spectacular star-forming trails behind ESO 137-001}",
      journal = {\mnras},
         year = 2023,
        month = jun,
       volume = {522},
       number = {1},
        pages = {173-194},
          doi = {10.1093/mnras/stad963},
archivePrefix = {arXiv},
       eprint = {2302.07270},
 primaryClass = {astro-ph.GA},
       adsurl = {https://ui.adsabs.harvard.edu/abs/2023MNRAS.522..173W}
}

@ARTICLE{Vollmer_2001,
       author = {{Vollmer}, B. and {Cayatte}, V. and {Balkowski}, C. and {Duschl}, W.~J.},
        title = "{Ram Pressure Stripping and Galaxy Orbits: The Case of the Virgo Cluster}",
      journal = {\apj},
         year = 2001,
        month = nov,
       volume = {561},
       number = {2},
        pages = {708-726},
          doi = {10.1086/323368},
archivePrefix = {arXiv},
       eprint = {astro-ph/0107237},
 primaryClass = {astro-ph},
       adsurl = {https://ui.adsabs.harvard.edu/abs/2001ApJ...561..708V}
}

@ARTICLE{Poggianti2025,
       author = {{Poggianti}, Bianca M. and {Vulcani}, Benedetta and {Tomicic}, Neven and {Moretti}, Alessia and {Gullieuszik}, Marco and {Bacchini}, Cecilia and {Fritz}, Jacopo and {George}, Koshy and {Gitti}, Myriam and {Ignesti}, Alessandro and {Jaff{\'e}}, Yara and {Lassen}, Augusto and {Marasco}, Antonino and {Radovich}, Mario and {Serra}, Paolo and {Smith}, Rory and {Tonnesen}, Stephanie and {Wolter}, Anna},
        title = "{The MUSE view of ram pressure stripped galaxies in clusters: The GASP sample}",
      journal = {\aap},
         year = 2025,
        month = jul,
       volume = {699},
          eid = {A357},
        pages = {A357},
          doi = {10.1051/0004-6361/202554200},
archivePrefix = {arXiv},
       eprint = {2505.21107},
 primaryClass = {astro-ph.GA},
       adsurl = {https://ui.adsabs.harvard.edu/abs/2025A&A...699A.357P}
}

@ARTICLE{Frank_1996,
       author = {{Frank}, Adam and {Jones}, T.~W. and {Ryu}, Dongsu and {Gaalaas}, Joseph B.},
        title = "{The Magnetohydrodynamic Kelvin-Helmholtz Instability: A Two-dimensional Numerical Study}",
      journal = {\apj},
         year = 1996,
        month = apr,
       volume = {460},
        pages = {777},
          doi = {10.1086/177009},
archivePrefix = {arXiv},
       eprint = {astro-ph/9510115},
 primaryClass = {astro-ph},
       adsurl = {https://ui.adsabs.harvard.edu/abs/1996ApJ...460..777F}
}

@ARTICLE{Salinas_2024,
       author = {{Salinas}, Vicente and {Jaff{\'e}}, Yara L. and {Smith}, Rory and {Shinn}, Jong-Ho and {Crossett}, Jacob P. and {Gullieuszik}, Marco and {Gonz{\'a}lez-Tor{\`a}}, Gemma and {Piraino-Cerda}, Franco and {Poggianti}, Bianca and {Vulcani}, Benedetta and {Biviano}, Andrea and {Louren{\c{c}}o}, Ana C.~C. and {Bilton}, Lawrence E. and {Kelkar}, Kshitija and {Calder{\'o}n-Castillo}, Paula},
        title = "{Constraining the duration of ram pressure stripping features in the optical from the direction of jellyfish galaxy tails}",
      journal = {\mnras},
         year = 2024,
        month = sep,
       volume = {533},
       number = {1},
        pages = {341-359},
          doi = {10.1093/mnras/stae1784},
archivePrefix = {arXiv},
       eprint = {2408.03396},
 primaryClass = {astro-ph.GA},
       adsurl = {https://ui.adsabs.harvard.edu/abs/2024MNRAS.533..341S}
}

@ARTICLE{Rintoul_2025,
       author = {{Rintoul}, Thomas A. and {van de Voort}, Freeke and {Hannington}, Andrew T. and {Pakmor}, R{\"u}diger and {Bieri}, Rebekka and {Werhahn}, Maria and {Talbot}, Rosie Y.},
        title = "{The role of magnetic fields in ram pressure stripping of satellite galaxies in the circumgalactic medium around massive galaxies}",
      journal = {arXiv e-prints},
         year = 2025,
        month = jun,
          eid = {arXiv:2506.18983},
        pages = {arXiv:2506.18983},
          doi = {10.48550/arXiv.2506.18983},
archivePrefix = {arXiv},
       eprint = {2506.18983},
 primaryClass = {astro-ph.GA},
       adsurl = {https://ui.adsabs.harvard.edu/abs/2025arXiv250618983R}
}

@ARTICLE{Girardi_1993,
       author = {{Girardi}, M. and {Biviano}, A. and {Giuricin}, G. and {Mardirossian}, F. and {Mezzetti}, M.},
        title = "{Velocity Dispersions in Galaxy Clusters}",
      journal = {\apj},
         year = 1993,
        month = feb,
       volume = {404},
        pages = {38},
          doi = {10.1086/172256},
       adsurl = {https://ui.adsabs.harvard.edu/abs/1993ApJ...404...38G}
}

@ARTICLE{McCourt_2015,
       author = {{McCourt}, Michael and {O'Leary}, Ryan M. and {Madigan}, Ann-Marie and {Quataert}, Eliot},
        title = "{Magnetized gas clouds can survive acceleration by a hot wind}",
      journal = {\mnras},
         year = 2015,
        month = may,
       volume = {449},
       number = {1},
        pages = {2-7},
          doi = {10.1093/mnras/stv355},
archivePrefix = {arXiv},
       eprint = {1409.6719},
 primaryClass = {astro-ph.GA},
       adsurl = {https://ui.adsabs.harvard.edu/abs/2015MNRAS.449....2M}
}

@ARTICLE{Sun_2022,
       author = {{Sun}, Ming and {Ge}, Chong and {Luo}, Rongxin and {Yagi}, Masafumi and {J{\'a}chym}, Pavel and {Boselli}, Alessandro and {Fossati}, Matteo and {Nulsen}, Paul E.~J. and {Yoshida}, Michitoshi and {Gavazzi}, Giuseppe},
        title = "{A universal correlation between warm and hot gas in the stripped tails of cluster galaxies}",
      journal = {Nature Astronomy},
         year = 2021,
        month = dec,
       volume = {6},
        pages = {270-274},
          doi = {10.1038/s41550-021-01516-8},
archivePrefix = {arXiv},
       eprint = {2103.09205},
 primaryClass = {astro-ph.GA},
       adsurl = {https://ui.adsabs.harvard.edu/abs/2022NatAs...6..270S}
}

@ARTICLE{Franchetto_2021,
       author = {{Franchetto}, Andrea and {Tonnesen}, Stephanie and {Poggianti}, Bianca M. and {Vulcani}, Benedetta and {Gullieuszik}, Marco and {Moretti}, Alessia and {Smith}, Rory and {Ignesti}, Alessandro and {Bacchini}, Cecilia and {McGee}, Sean and {Tomi{\v{c}}i{\'c}}, Neven and {Mingozzi}, Matilde and {Wolter}, Anna and {M{\"u}ller}, Ancla},
        title = "{Evidence for Mixing between ICM and Stripped ISM by the Analysis of the Gas Metallicity in the Tails of Jellyfish Galaxies}",
      journal = {\apjl},
         year = 2021,
        month = nov,
       volume = {922},
       number = {1},
          eid = {L6},
        pages = {L6},
          doi = {10.3847/2041-8213/ac3664},
archivePrefix = {arXiv},
       eprint = {2111.04755},
 primaryClass = {astro-ph.GA},
       adsurl = {https://ui.adsabs.harvard.edu/abs/2021ApJ...922L...6F}
}

@ARTICLE{Ignesti_2024,
       author = {{Ignesti}, Alessandro and {Brunetti}, Gianfranco and {Gullieuszik}, Marco and {Akerman}, Nina and {Marasco}, Antonino and {Poggianti}, Bianca M. and {Li}, Yuan and {Vulcani}, Benedetta and {Gitti}, Myriam and {Moretti}, Alessia and {Giunchi}, Eric and {Tomi{\v{c}}i{\'c}}, Neven and {Bacchini}, Cecilia and {Paladino}, Rosita and {Radovich}, Mario and {Wolter}, Anna},
        title = "{Investigating the Intracluster Medium Viscosity Using the Tails of GASP Jellyfish Galaxies}",
      journal = {\apj},
         year = 2024,
        month = dec,
       volume = {977},
       number = {2},
          eid = {219},
        pages = {219},
          doi = {10.3847/1538-4357/ad919b},
archivePrefix = {arXiv},
       eprint = {2411.07034},
 primaryClass = {astro-ph.CO},
       adsurl = {https://ui.adsabs.harvard.edu/abs/2024ApJ...977..219I}
}

@ARTICLE{Jaffe_2018,
       author = {{Jaff{\'e}}, Yara L. and {Poggianti}, Bianca M. and {Moretti}, Alessia and {Gullieuszik}, Marco and {Smith}, Rory and {Vulcani}, Benedetta and {Fasano}, Giovanni and {Fritz}, Jacopo and {Tonnesen}, Stephanie and {Bettoni}, Daniela and {Hau}, George and {Biviano}, Andrea and {Bellhouse}, Callum and {McGee}, Sean},
        title = "{GASP. IX. Jellyfish galaxies in phase-space: an orbital study of intense ram-pressure stripping in clusters}",
      journal = {\mnras},
         year = 2018,
        month = jun,
       volume = {476},
       number = {4},
        pages = {4753-4764},
          doi = {10.1093/mnras/sty500},
archivePrefix = {arXiv},
       eprint = {1802.07297},
 primaryClass = {astro-ph.GA},
       adsurl = {https://ui.adsabs.harvard.edu/abs/2018MNRAS.476.4753J}
}

@ARTICLE{Rohr_2023,
       author = {{Rohr}, Eric and {Pillepich}, Annalisa and {Nelson}, Dylan and {Zinger}, Elad and {Joshi}, Gandhali D. and {Ayromlou}, Mohammadreza},
        title = "{Jellyfish galaxies with the IllustrisTNG simulations - when, where, and for how long does ram pressure stripping of cold gas occur?}",
      journal = {\mnras},
         year = 2023,
        month = sep,
       volume = {524},
       number = {3},
        pages = {3502-3525},
          doi = {10.1093/mnras/stad2101},
archivePrefix = {arXiv},
       eprint = {2304.09196},
 primaryClass = {astro-ph.GA},
       adsurl = {https://ui.adsabs.harvard.edu/abs/2023MNRAS.524.3502R}
}

@ARTICLE{Wetzel_2010,
       author = {{Wetzel}, Andrew R. and {White}, Martin},
        title = "{What determines satellite galaxy disruption?}",
      journal = {\mnras},
         year = 2010,
        month = apr,
       volume = {403},
       number = {2},
        pages = {1072-1088},
          doi = {10.1111/j.1365-2966.2009.16191.x},
archivePrefix = {arXiv},
       eprint = {0907.0702},
 primaryClass = {astro-ph.CO},
       adsurl = {https://ui.adsabs.harvard.edu/abs/2010MNRAS.403.1072W}
}

@ARTICLE{Rhee_2017,
       author = {{Rhee}, Jinsu and {Smith}, Rory and {Choi}, Hoseung and {Yi}, Sukyoung K. and {Jaff{\'e}}, Yara and {Candlish}, Graeme and {S{\'a}nchez-J{\'a}nssen}, Ruben},
        title = "{Phase-space Analysis in the Group and Cluster Environment: Time Since Infall and Tidal Mass Loss}",
      journal = {\apj},
         year = 2017,
        month = jul,
       volume = {843},
       number = {2},
          eid = {128},
        pages = {128},
          doi = {10.3847/1538-4357/aa6d6c},
archivePrefix = {arXiv},
       eprint = {1704.04243},
 primaryClass = {astro-ph.GA},
       adsurl = {https://ui.adsabs.harvard.edu/abs/2017ApJ...843..128R}
}

@ARTICLE{Armillotta_2017,
       author = {{Armillotta}, L. and {Fraternali}, F. and {Werk}, J.~K. and {Prochaska}, J.~X. and {Marinacci}, F.},
        title = "{The survival of gas clouds in the circumgalactic medium of Milky Way-like galaxies}",
      journal = {\mnras},
         year = 2017,
        month = sep,
       volume = {470},
       number = {1},
        pages = {114-125},
          doi = {10.1093/mnras/stx1239},
archivePrefix = {arXiv},
       eprint = {1608.05416},
 primaryClass = {astro-ph.GA},
       adsurl = {https://ui.adsabs.harvard.edu/abs/2017MNRAS.470..114A}
}

@ARTICLE{Pratt_2022,
       author = {{Pratt}, G.~W. and {Arnaud}, M. and {Maughan}, B.~J. and {Melin}, J. -B.},
        title = "{Linking a universal gas density profile to the core-excised X-ray luminosity in galaxy clusters up to z {\ensuremath{\sim}} 1.1}",
      journal = {\aap},
         year = 2022,
        month = sep,
       volume = {665},
          eid = {A24},
        pages = {A24},
          doi = {10.1051/0004-6361/202243074},
archivePrefix = {arXiv},
       eprint = {2206.06656},
 primaryClass = {astro-ph.CO},
       adsurl = {https://ui.adsabs.harvard.edu/abs/2022A&A...665A..24P}
}

@ARTICLE{Wez_2011,
       author = {{We{\.z}gowiec}, M. and {Vollmer}, B. and {Ehle}, M. and {Dettmar}, R. -J. and {Bomans}, D.~J. and {Chy{\.z}y}, K.~T. and {Urbanik}, M. and {Soida}, M.},
        title = "{Hot gas in Mach cones around Virgo cluster spiral galaxies}",
      journal = {\aap},
         year = 2011,
        month = jul,
       volume = {531},
          eid = {A44},
        pages = {A44},
          doi = {10.1051/0004-6361/201016344},
archivePrefix = {arXiv},
       eprint = {1104.2713},
 primaryClass = {astro-ph.CO},
       adsurl = {https://ui.adsabs.harvard.edu/abs/2011A&A...531A..44W}
}

@ARTICLE{George_2025,
       author = {{George}, K. and {Poggianti}, B.~M. and {Vulcani}, B. and {Gullieuszik}, M. and {Postma}, J. and {Fritz}, J. and {C{\^o}t{\'e}}, P. and {Jaffe}, Y.~L. and {Moretti}, A. and {Ignesti}, A. and {Peluso}, G. and {Tomi{\'c}i{\'c}}, N. and {Subramaniam}, A. and {Ghosh}, S.~K. and {Tandon}, S.~N.},
        title = "{Star formation at different stages of ram-pressure stripping as observed through far-ultraviolet imaging of 13 GASP galaxies}",
      journal = {\aap},
         year = 2025,
        month = aug,
       volume = {700},
          eid = {A38},
        pages = {A38},
          doi = {10.1051/0004-6361/202554945},
archivePrefix = {arXiv},
       eprint = {2505.15066},
 primaryClass = {astro-ph.GA},
       adsurl = {https://ui.adsabs.harvard.edu/abs/2025A&A...700A..38G}
}

@ARTICLE{Westmeier2021MNRAS.506.3962W,
       author = {{Westmeier}, T. and {Kitaeff}, S. and {Pallot}, D. and {Serra}, P. and {van der Hulst}, J.~M. and {Jurek}, R.~J. and {Elagali}, A. and {For}, B. -Q. and {Kleiner}, D. and {Koribalski}, B.~S. and {Lee-Waddell}, K. and {Mould}, J.~R. and {Reynolds}, T.~N. and {Rhee}, J. and {Staveley-Smith}, L.},
        title = "{SOFIA 2 - an automated, parallel H I source finding pipeline for the WALLABY survey}",
      journal = {\mnras},
         year = 2021,
        month = sep,
       volume = {506},
       number = {3},
        pages = {3962-3976},
          doi = {10.1093/mnras/stab1881},
archivePrefix = {arXiv},
       eprint = {2106.15789},
 primaryClass = {astro-ph.IM},
       adsurl = {https://ui.adsabs.harvard.edu/abs/2021MNRAS.506.3962W}
}

@ARTICLE{huts,
       author = {{Hutschenreuter}, S. and {Anderson}, C.~S. and {Betti}, S. and {Bower}, G.~C. and {Brown}, J.-A. and {Br{\"u}ggen}, M. and {Carretti}, E. and {Clarke}, T. and {Clegg}, A. and {Costa}, A. and {Croft}, S. and {Van Eck}, C. and {Gaensler}, B.~M. and {de Gasperin}, F. and {Haverkorn}, M. and {Heald}, G. and {Hull}, C.~L.~H. and {Inoue}, M. and {Johnston-Hollitt}, M. and {Kaczmarek}, J. and {Law}, C. and {Ma}, Y.~K. and {MacMahon}, D. and {Mao}, S.~A. and {Riseley}, C. and {Roy}, S. and {Shanahan}, R. and {Shimwell}, T. and {Stil}, J. and {Sobey}, C. and {O'Sullivan}, S.~P. and {Tasse}, C. and {Vacca}, V. and {Vernstrom}, T. and {Williams}, P.~K.~G. and {Wright}, M. and {En{\ss}lin}, T.~A.},
        title = "{The Galactic Faraday rotation sky 2020}",
      journal = {\aap},
         year = 2022,
        month = jan,
       volume = {657},
          eid = {A43},
        pages = {A43},
          doi = {10.1051/0004-6361/202140486},
archivePrefix = {arXiv},
       eprint = {2102.01709},
 primaryClass = {astro-ph.GA},
       adsurl = {https://ui.adsabs.harvard.edu/abs/2022A&A...657A..43H}
}





\end{document}